\documentclass[twocolumn,12pt]{aastex62}

\usepackage{epstopdf}
\usepackage{amsmath}
\usepackage[T1]{fontenc}

\newcommand{\cucm}{\ensuremath{\textrm{ cm}^{-3}}}

\newcommand{\kpc}{\ensuremath{\, \mathrm{kpc}}}
\newcommand{\pc}{\ensuremath{\textrm{ pc}}}
\newcommand{\K}{\ensuremath{\textrm{ K}}}
\newcommand{\erg}{\ensuremath{\textrm{ erg}}}
\newcommand{\s}{\ensuremath{\textrm{ s}}}
\newcommand{\uG}{\ensuremath{\, \mu \textrm{G}}}

\newcommand{\Myr}{\ensuremath{\textrm{ Myr}}}
\newcommand{\kyr}{\ensuremath{\textrm{ kyr}}}

\newcommand{\Pmin}{\ensuremath{P_\mathrm{min}}}
\newcommand{\Pmax}{\ensuremath{P_\mathrm{max}}}
\newcommand{\Pth}{\ensuremath{P_\mathrm{th}}}

\newcommand{\Pturb}{\ensuremath{P_\mathrm{turb}}}
\newcommand{\Ptotz}{\ensuremath{P_{\mathrm{tot,}z}}}
\newcommand{\Ptot}{\ensuremath{P_\mathrm{tot}}}
\newcommand{\GPE}{\ensuremath{\Gamma_{\mathrm{pe}}}}
\newcommand{\nbarwc}{\ensuremath{\langle n \rangle_\mathrm{w,c}}}

\renewcommand{\vec}{\mathbf}

\shorttitle{Stability of the three-phase ISM}
\shortauthors{Hill, Mac Low, Gatto, \& Ib\'a\~nez-Mej\'ia}

\received{April 12, 2018}
\revised{June 12, 2018}
\accepted{June 12, 2018}
\journalinfo{The Astrophysical Journal in press}

\begin{document}

\date{\today}

\title{Effect of the heating rate on the stability of the three-phase interstellar medium}
\author[0000-0001-7301-5666]{Alex S. Hill}
\affiliation{Departments of Physics and Astronomy, Haverford College, Haverford, PA 19041 USA}
\affiliation{Department of Physics and Astronomy, University of British Columbia, Vancouver, BC V6T 1Z1 Canada}
\affiliation{Space Science Institute, Boulder, CO 80301 USA}
\affiliation{Dominion Radio Astrophysical Observatory, Herzberg Program in Astronomy and Astrophysics, National Research Council Canada, Penticton, BC V2A 6J9 Canada}
\author[0000-0003-0064-4060]{Mordecai-Mark Mac Low}
\affiliation{Department of Astrophysics, American Museum of Natural History, New York, NY 10024 USA}
\author[0000-0002-6838-5248]{Andrea Gatto}
\affiliation{Space Dynamics Services, via Mario Giuntini 63, I-56023 Cascina, Italy}
\author[0000-0002-9868-3561]{Juan C. Ib\'a\~nez-Mej\'ia}
\affiliation{I. Physikalisches Institut, Universit\"at zu K\"oln, D-50937 K\"oln, Germany}
\email{ashill@astro.ubc.ca}
\begin{abstract}
We investigate the impact of the far ultraviolet (FUV) heating rate on the stability of the three-phase interstellar medium using three-dimensional simulations of a $1\kpc^2$, vertically-extended domain. The FUV heating rate sets the range of thermal pressures across which the cold ($\sim10^2\K$) and warm ($\sim10^4\K$) neutral media (CNM and WNM) can coexist in equilibrium. 
Even absent a variable star formation rate regulating the FUV heating rate, the gas physics keeps the pressure in the two-phase regime: because radiative heating and cooling processes happen on shorter timescales than sound wave propagation, turbulent compressions tend to keep the interstellar medium within the CNM-WNM pressure regime over a wide range of heating rates. The thermal pressure is set primarily by the heating rate with little influence from the hydrostatics. The vertical velocity dispersion adjusts as needed to provide hydrostatic support given the thermal pressure: when the turbulent pressure $\langle\rho\rangle\sigma_z^2$ is calculated over scales $\gtrsim500\pc$, the thermal plus turbulent pressure approximately equals the weight of the gas.
The warm gas volume filling fraction is $0.2<f_w<0.8$ over a factor of less than three in heating rate, with $f_w$ near unity at higher heating rates and near zero at lower heating rates.
We suggest that cosmological simulations that do not resolve the CNM should maintain an interstellar thermal pressure within the two-phase regime.
 \end{abstract}


\section{Introduction}

Normal (non-starburst) star-forming galaxies generally have thermally stable gas in their interstellar media (ISM) at both cold ($T \lesssim 300 \K$) and warm ($T \sim 5000-12000 \K$) temperatures \citep[see reviews by][]{Ferriere:2001vr,Cox:2005}. The cold phase consists of molecular gas --- where nearly all star formation occurs --- and cold atomic gas called the cold neutral medium (CNM). The warm phase consists of both neutral gas (the warm neutral medium, WNM) and photoionized gas (the warm ionized medium, WIM).

Theoretically, there is a relatively narrow range of thermal pressures in which the CNM and WNM can coexist in thermal pressure equilibrium for a given heating rate and cooling function \citep{FieldGoldsmith:1969,WolfireHollenbach:1995,WolfireMcKee:2003}. Observationally, nearly all normal star-forming galaxies have a WIM \citep{OeyMeurer:2007}. The \ion{H}{1} 21~cm emission to absorption ratio in the Milky Way is constant out to a Galactocentric radius of at least $25 \kpc$, indicating that the CNM-WNM density ratio is constant even in the outer Milky Way where the pressure is most likely low \citep{DickeyStrasser:2009}. Only $\approx 0.05\%$ of the CNM has a pressure above the two-phase range \citep{JenkinsTripp:2001,JenkinsTripp:2011}. These observations suggest that cold and warm gas coexist in most places in most star-forming galaxies, indicating that the gas pressure must be within the stable two-phase regime. \citet{WolfireMcKee:2003} argue that turbulent compression combined with the thermal instability physics drives the thermal pressure into the stable two-phase range.

In addition to the cold and warm phases of the ISM, shock heating due to supernova explosions creates a third quasi-stable phase of hot ($\gtrsim 10^6 \K$) gas called the hot ionized medium (HIM), identified observationally from \ion{O}{6} absorption lines \citep{JenkinsMeloy:1974,York:1974} and the soft X-ray background \citep[eg][]{Burstein:1977,McCammonSanders:1990,McCammonAlmy:2002,HenleyShelton:2010a,HenleyShelton:2013,HenleyShelton:2010,HenleyShelton:2015}. 

Because the density of the HIM is very low, the cooling time is long compared to the typical time between passing shocks, making the HIM long-lived even though it is not in equilibrium.
\citet{CoxSmith:1974} defined the porosity, $q$, such that the fraction of the volume occupied by the HIM is $f_h = 1-\exp(-q)$. In nearly all cases, the cold gas occupies a small fraction of the volume, so the warm gas filling factor is $f_w \approx 1 - f_h$.
\citet{McKeeOstriker:1977} combined the two-phase model with calculations of the effect of hot gas to describe a three-phase ISM regulated by supernovae. They found that the HIM dominates, with $q \gtrsim 3$, and the CNM, WNM, and WIM are confined to clouds embedded in the HIM. This is a thermal runaway state: because of the very long cooling time of the HIM ($t_\mathrm{cool} \sim kT/n^2 \Lambda \sim$ several Gyr for $T \gtrsim 5 \times 10^6 \K$ gas in pressure equilibrium with the WNM), it is difficult to form warm gas once the volume is dominated by hot gas. Therefore, subsequent supernovae continue to heat the gas; the excess energy is not radiated away. In the McKee-Ostriker picture, the HIM contains very little mass but has sufficient thermal pressure to confine the cold-warm clouds.

However, due to the gravity of the Galaxy, overpressured, low-density hot gas tends to buoyantly move away from the midplane, reducing the pressure confining the WNM and CNM and driving a cycle of matter and energy between the disk and the halo called a Galactic fountain \citep{Bregman:1980gn,RosenBregman:1995,WaltersCox:2001}. Although nonthermal sources of pressure dominate the vertical support of the ISM \citep{BoularesCox:1990,GirichidisWalch:2016}, this can act as a pressure valve. Because the magnetic field is very weak in the hot gas, it acts as an additional pressure term in the warm and cold gas, compressing the hot gas \citep{SlavinCox:1992,SlavinCox:1993}. Modern hydrodynamic models of the ISM in normal galaxies which incorporate vertical gravity, domains large enough in the vertical direction to track fountain flows, energy injection due to supernovae and resulting turbulent mixing, diffuse heating due to FUV radiation, and optically-thin line cooling in three dimensions have generally found that the hot gas is confined to a minority of the midplane volume, $q \sim 0.3$ \citep{KorpiBrandenburg:1999,Avillez:2000,AvillezBreitschwerdt:2004,JoungMac-Low:2006,HillJoung:2012,GentShukurov:2013a}. This result holds for a wide variety of gas surface mass densities when the star formation rate is adjusted following a Kennicutt-Schmidt relation \citep{AvillezBreitschwerdt:2004,JoungMac-Low:2009}. Across the cold, warm, and hot phases of the ISM, the gas is isobaric in total pressure, with the turbulent pressure increasingly-dominant at high star formation rates \citep{JoungMac-Low:2009}. However, \citet{JoungMac-Low:2009} needed to tune their star formation rate carefully in order to maintain the three-phase medium.

The most recent numerical simulations have tended to again invert this picture. \citet{WalchGirichidis:2015} determine supernova times randomly but use a variety of fractions of ``peak driving'', in which the supernovae are set off at density peaks. They find that, for many supernova rates with random or a mix of random and peak driving, the simulations are in a thermal runaway state, with $q \gtrsim 2$. Pure peak driving leads to a loss of most of the injected thermal energy due to the high gas density and thus high cooling rate. In these models, mixed driving is needed for a three-phase ISM to exist, at least in models with fully periodic boundary conditions \citep{GattoWalch:2015}. \citet{LiOstriker:2015} argued, also using fully-periodic boundary conditions, that the photoelectric heating rate controls the porosity, with higher heating rates leading to higher warm gas thermal pressures which are more effective at confining supernova remnants.

\citet{OstrikerMcKee:2010} and \citet{OstrikerShetty:2011} developed a model based on the assumption that star-forming galaxies maintain both thermal and vertical dynamical equilibrium. In their picture, the thermal pressure is stable to perturbation out of the two-phase regime because the star formation rate adjusts in order to maintain these dual equilibria; galaxies self-regulate their star formation rates.

In this work, we investigate the impact of the gas physics on the thermal pressure without a star formation rate that responds to the gas density and the ensuing self-regulation. We vary photoelectric heating rates and explore the porosity of the ISM in simulations of supernova-driven turbulence with vertical stratification. We describe our simulations in Section~\ref{sec:simulations}, discuss the CNM-WNM two-phase equilibrium in Section~\ref{sec:CNMWNM}, and discuss the hydrostatic support of the ISM in Section~\ref{sec:hydrostatic}. We describe the characteristics of our simulations in Section~\ref{sec:results}. We then discuss the implications in Section~\ref{sec:discussion}, presenting a physical argument explaining the stability of the three-phase ISM in Section~\ref{sec:stability} and discussing the limitations of our physical approximations in Section~\ref{sec:missing}. We end with a summary of our main points in Section~\ref{sec:summary}.

\section{Simulations} \label{sec:simulations}

\begin{deluxetable*}{l r@{ -- }r ccc D D D}
\tablecaption{Simulations used in this paper
\label{tbl:runs}}
\tabletypesize{\footnotesize}
\decimals
\tablehead{\colhead{Name} & \twocolhead{$t$ range} & \colhead{$B_{x,0}$} & \colhead{$\log n_\mathrm{IGM}$} & \colhead{$\Sigma_{\mathrm{gas}}$} & \twocolhead{$\Delta x_0$}  & \twocolhead{SN rate} & \twocolhead{$\Gamma_0/10^{-25}$}  \\
 & \twocolhead{(Myr)} & \colhead{($\mu \mathrm{G}$)} & \colhead{($\mathrm{cm}^{-3}$)} & \colhead{($M_{\odot} \, \pc^{-2}$)} &  \twocolhead{(pc)} & \twocolhead{(Myr$^{-1}$ kpc$^{-2}$)} & \twocolhead{(erg s$^{-1}$)} }
\colnumbers
\startdata
\multicolumn{12}{c}{Fiducial} \\
\hline
{\tt pe1229}   			&   0 & 160 & 0   & $-7.1$ & 13.2   & 7.81 & 34.1 & 12.30   \\
{\tt pe109}          	& 160 & 200 &\nodata & \nodata & \nodata   & 3.91 & 34.1 & 1.09   \\
{\tt pe109 2pc}        	& 190 & 200 &\nodata & \nodata & \nodata   & 1.95 & 34.1 & 1.09   \\
{\tt pe200}          	& 160 & 200 &\nodata & \nodata & \nodata   & 3.91 & 34.1 & 2.00   \\
{\tt pe200 2pc}         & 190 & 200 &\nodata & \nodata & \nodata   & 1.95 & 34.1 & 2.00   \\
{\tt pe300}          	& 160 & 360 &\nodata & \nodata & \nodata   & 3.91 & 34.1 & 3.00   \\
{\tt pe300 2pc}         & 190 & 200 &\nodata & \nodata & \nodata   & 1.95 & 34.1 & 3.00   \\
{\tt sn17 pe300}        & 160 & 200 &\nodata & \nodata & \nodata   & 3.91 & 17.1 & 3.00   \\
{\tt pe400}          	& 160 & 200 &\nodata & \nodata & \nodata   & 3.91 & 34.1 & 4.00   \\
{\tt pe400 2pc}         & 190 & 200 &\nodata & \nodata & \nodata   & 1.95 & 34.1 & 4.00   \\
{\tt pe500}          	& 160 & 200 &\nodata & \nodata & \nodata   & 3.91 & 34.1 & 5.00   \\
{\tt pe500 2pc}         & 190 & 200 &\nodata & \nodata & \nodata   & 1.95 & 34.1 & 5.00   \\
{\tt pe600}          	& 160 & 200 &\nodata & \nodata & \nodata   & 3.91 & 34.1 & 6.00   \\
{\tt pe600 2pc}         & 190 & 200 &\nodata & \nodata & \nodata   & 1.95 & 34.1 & 6.00   \\
{\tt pe1229}	      	& 160 & 200 &\nodata & \nodata & \nodata   & 3.91 & 34.1 & 12.30   \\
{\tt pe1229 2pc}      	& 190 & 200 &\nodata & \nodata & \nodata   & 1.95 & 34.1 & 12.30   \\
{\tt pe1229 1pc}        & 190 & 200 &\nodata & \nodata & \nodata   & 0.98 & 34.1 & 12.30   \\
\hline
\multicolumn{12}{c}{Magnetized} \\
\hline
{\tt bx5 pe300}      	& 0 & 160 & 5   & $-7.1$ & 13.2   & 7.81 & 34.1 & 3.00   \\
{\tt bx5 pe300}      	& 160 & 360 &\nodata & \nodata & \nodata   & 3.91 & 34.1 & 3.00   \\
{\tt bx5 pe1229}     	& 160 & 200 &\nodata & \nodata & \nodata   & 3.91 & 34.1 & 12.30   \\
\hline
\multicolumn{12}{c}{$n_\mathrm{IGM} = 10^{-4.1} \cucm$, unmagnetized} \\
\hline
{\tt nigm-4 bx0 pe300}  & 0 & 160 & 0   & $-4.1$ & 13.3   & 7.81 & 34.1 & 3.00   \\
{\tt nigm-4 bx0 pe300}  & 160 & 200 &\nodata & \nodata & \nodata   & 3.91 & 34.1 & 3.00   \\
\hline
\multicolumn{12}{c}{$n_\mathrm{IGM} = 10^{-4.1} \cucm$, magnetized} \\
\hline
{\tt nigm-4 bx5 pe300}  & 0 & 160 & 5   & $-4.1$ & 13.4   & 7.81 & 34.1 & 3.00   \\
{\tt nigm-4 bx5 pe300}  & 160 & 200 &\nodata & \nodata & \nodata   & 3.91 & 34.1 & 3.00   \\
\hline
\multicolumn{12}{c}{$\Sigma_\mathrm{gas} = 7.5 \, M_\odot \pc^{-2}$} \\
\hline
{\tt smd75 pe300}   	& 0 & 160 & 0   & $-7.1$ & 7.5   & 7.81 & 34.1  & 3.00   \\
{\tt smd75 pe300}   	& 160 & 200 &\nodata & \nodata & \nodata   & 3.91 & 34.1  & 3.00   \\
{\tt smd75 pe600}   	& 160 & 200 &\nodata & \nodata & \nodata   & 3.91 & 34.1  & 6.00   \\
\enddata
\tablecomments{Runs used in this paper. Base runs with $8 \pc$ resolution are listed first in each section. Science runs, with runtime parameters modified and $1-4 \pc$ resolution, are resumed at $t=160 \Myr$ from the corresponding base run.}
\end{deluxetable*}

We use simulations in which a vertically-stratified, multiphase ISM is established by including a fixed cooling curve, a diffuse heating term modeling photoelectric heating of dust grains by the stellar FUV radiation field, supernova explosions, a fixed gravitational potential, and magnetic fields. We originally implemented our model using Flash 2.5 \citep{FryxellOlson:2000} to solve the magnetohydrodynamic (MHD) equations numerically \citep{JoungMac-Low:2006,JoungMac-Low:2009,HillJoung:2012}. In this work, we use a reimplementation of the model in Flash 4.2 \citep{Baczynski:2016,Ibanez-MejiaMac-Low:2016}. We refer the reader to \citet{HillJoung:2012} and references therein for a complete description of the model.

We use a $1 \times 1 \times \pm 20 \kpc^3$ simulation domain with periodic boundary conditions on the sides and outflow boundary conditions on the top and bottom. Using adaptive mesh refinement, the grid is fully refined ($\Delta x_0 = 1-8 \pc$) near the midplane with decreasing resolution with height; the lowest resolution is $\Delta x = 32 \pc$. We use a fixed gravitational potential \citep{KuijkenGilmore:1989,NavarroFrenk:1996}. In the initial condition, we establish isothermal gas in hydrostatic equilibrium with a density floor $n_\mathrm{IGM}$ added. We specify the cooling function $\Lambda(T)$ for solar metallicity gas with no depletion with an ionization fraction of $10^{-2}$ in the atomic gas (\citealt{JoungMac-Low:2006}, based on \citealt{DalgarnoMcCray:1972,SutherlandDopita:1993}); in this approximation, the range of thermal instability is $39.8 \K < T < 10^4 \K$ with $39.8 \K < T < 200 \K$ marginally stable.\footnote{We define ``cold'' gas as $T < 39.8 \K$, ``cool'' as $39.8 \K < T < 10^4 \K$, ``warm'' as $10^4 \K < T < 2 \times 10^4 \K$, ``transition-temperature'' as $2 \times 10^4 \K < T < 10^{5.5} \K$, and ``hot'' as $T > 10^{5.5} \K$.} We apply a diffuse heating term, which manifests itself in the energy equation for energy density $E$ as
\begin{equation}
\frac{\partial E}{\partial t} = n \GPE(z,T) - n^2 \Lambda(T).
\end{equation}
The heating term is given by
\begin{equation} \label{eq:heating}
\GPE(z,T) = \left\{ \begin{array}{rl}
\Gamma_0 e^{-|z| / h_\mathrm{pe}}, & T < 2 \times 10^4 \K \\
 0, & T > 2 \times 10^4 \K.
\end{array} \right.
\end{equation}
This heating term is designed to approximate photoelectric heating of dust grains. In our model, equation~(\ref{eq:heating}) along with supernovae and stellar winds give the only heating source terms. We set $h_\mathrm{pe} = 8.5 \kpc$ (compared to $0.3 \kpc$ in our previous work), so the heating rate is roughly constant up to large heights. Although the scale height of the thin stellar disk is only $\approx 100 \pc$ \citep[e.g.][]{KongZhu:2008}, the FUV radiation field falls off much more slowly with height because surface brightness is conserved. This is supported observationally by the presence of a two-phase ISM in high velocity clouds \citep[e.g.][]{MossMcClure-Griffiths:2013}; in models with $h_\mathrm{pe} = 0.3 \kpc$, high velocity cloud-like regions entirely cool to CNM temperatures, which is inconsistent with observations. An estimation of the FUV flux of M81 also suggests that the FUV flux decreases slowly with height \citep{HillBohlin:1992,WolfireMcKee:1995}. Other groups constructing similar simulations have used a FUV radiation field which does not vary at all with height \citep{WalchGirichidis:2015,KimOstriker:2017a}. This choice ultimately has little impact on the midplane behavior that is the main focus of this paper.

Because the FUV radiation field in the diffuse ISM varies widely and because we do not account for other heating terms, we have considerable freedom to vary $\Gamma_0$; doing so is the main focus of this paper.  For typical photoelectric heating with a heating efficiency $\epsilon \approx 0.5$ and a FUV radiation field $G_0=1.7$ times the interstellar field near the Sun \citep{Habing:1968uc}, the midplane heating rate$\Gamma_0 = 0.85 \times 10^{-25} \erg \s^{-1}$, the value used in \citet{JoungMac-Low:2006} and \citet{HillJoung:2012}.

We set off supernovae in three distinct populations as in \citet{JoungMac-Low:2006}: Type Ia supernovae ($19\%$ of all supernovae), which are distributed exponentially with scale height $300 \pc$; field core-collapse supernovae ($32\%$), which are distributed exponentially with scale height $90 \pc$; and correlated core-collapse supernovae ($47\%$), which also have an exponential distribution with scale height $90 \pc$. For the correlated supernoave, between $7$ and $40$ supernovae are set off at the same position, spread evenly over $40 \Myr$, to simulate an OB association. For the first $5 \Myr$, we continuously inject $10^{51} \erg$ (i.e.\ $6.3 \times 10^{36} \textrm{ erg} \s^{-1}$) of thermal energy in a single zone to approximate the clearing effect of winds from massive stars. For each supernova, we inject $10^{51} \erg$ of thermal energy in one timestep in a $60 \, M_\odot$ sphere and redistribute the mass evenly around the sphere; we do not inject momentum.

In all cases, we choose the location and timing of supernovae {\em without knowledge of the gas density}. This is somewhat unphysical, although we do not resolve star formation in any meaningful way in these models and the B stars that produce most supernovae even in massive star clusters live longer than the giant molecular clouds in which they form. The major drawback of this piece of missing physics is that cold clouds are unphysically long-lived (with individual cold clouds surviving for $\gtrsim 100 \Myr$) due to the lack of internal feedback. This setup allows us to conduct a numerical physics experiment to study the effects of the heating rate independent of changes in the supernova history. In models with a given supernova rate, we chose the random seeds such that the supernova history (time and position) is identical, enabling direct comparison of runs with different heating rates.

We include two models with a magnetic field with initial conditions the same as \citet{HillJoung:2012}: a horizontal field with midplane strength $B_{x,0} = 5 \uG$, decreasing as a function of height to maintain a constant plasma $\beta$.

The runs we consider in this paper are listed in Table~\ref{tbl:runs}. We used a set of low-resolution ($\Delta x_0 = 8 \pc$) runs with a range of initial conditions of the surface mass density $\Sigma_\mathrm{gas}$, density floor $n_\mathrm{IGM}$, and magnetic field to establish turbulent initial conditions, running these base runs for $160 \Myr$. We then switched to higher resolution, $\Delta x_0 = 4 \pc$, and resumed the run; in some cases, we ran from $190-200 \Myr$ with $\Delta x_0 = 2 \pc$. In the highest heating rate model, we also ran a $\Delta x_0 = 1 \pc$ run (see Section~\ref{sec:CNMWNM} below). For these resumed runs, we varied runtime inputs which do not require restarting from scratch: the diffuse heating rate, the supernova rate, and the resolution. We typically consider snapshots after $200 \Myr$ of evolution, $40 \Myr$ after switching to the final resolution and heating rate. This is sufficient time for the volume filling fractions in the midplane to reach a dynamical equilibrium \citep{AvillezBreitschwerdt:2004,HillJoung:2012}. We ran two models to $360 \Myr$ to verify that the models reach a dynamical equilibrium by $200 \Myr$.

We discuss the impact of a number of neglected physical processes in Section~\ref{sec:missing} below.

\section{Coexistence of the CNM and WNM} \label{sec:CNMWNM}

The physical conditions in the CNM and WNM are determined primarily by the balance between photoelectric heating, parameterized by \GPE, and radiative cooling, parameterized by the cooling function $\Lambda(T)$. Neglecting other effects, the ISM at number density $n$ is in thermal equilibrium when heating equals cooling,
\begin{equation} \label{eq:equilib}
n \GPE = n^2 \Lambda(T).
\end{equation}
The shape of $\Lambda(T)$ yields a narrow range of thermal pressures in which equation~(\ref{eq:equilib}) can be satisfied for both the CNM and the WNM, the pressure regime in which a two-phase atomic medium can exist. We denote this pressure range as $\Pmin < \Pth < \Pmax$.
If $\Pth > \Pmax$, one expects that the CNM is the dominant phase by mass, whereas the WNM is expected to be the dominant phase by mass if $\Pth < \Pmin$. If WNM gas exceeds \Pmax, the heating is insufficient to prevent gas from cooling to CNM temperatures. If the thermal pressure in cold gas drops below \Pmin, the cooling is insufficient to prevent gas from heating to WNM temperatures. For a given $\Lambda$ (determined by the given metallicity and ionization state and included cooling processes such as non-equilibrium effects), the values of \Pmin\ and \Pmax\ are set by the heating rate \GPE. This is in reality typically a weak function of gas temperature; in our model, it is independent of gas temperature for $T \le 2 \times 10^4 \K$ and vanishes at higher temperatures. 
The FUV radiation field that controls \GPE, however, comes primarily from short-lived stars and thus changes by factors of $\sim 3$ on $\lesssim 100 \Myr$ timescales in the ISM near the Sun \citep{ParravanoHollenbach:2003}. For our cooling curve, the minimum stable CNM thermal pressure and maximum stable WNM thermal pressure are
\begin{equation} \label{eq:Pminmax}
(\Pmin, \Pmax) = (665.2, 8269.4) \, \Gamma_{-25} \, k \K \cucm,
\end{equation}
where the subscript $-25$ denotes $\Gamma \cdot (10^{-25} \erg \s^{-1})^{-1}$. This range is relatively wide, spanning more than a decade. For the cooling curve used by \citet{WolfireMcKee:2003}, for example, $2000 \K \cucm < \Pth/k < 5000 \K \cucm$ for the solar neighborhood.

The typical temperature of the cold gas does not change as the heating rate increases. Therefore, the density of CNM gas at \Pmin\ is
\begin{equation} \label{eq:nCNMmin}
n_\mathrm{CNM,min} = \frac{\Pmin}{kT} = 33.3 \, \Gamma_{-25} \cdot \left( \frac{T}{20 \K} \right)^{-1} \cucm.
\end{equation}
Because the density of cold gas increases with the heating rate, the numerical resolution required to resolve typical cold clumps also increases with the heating rate. The linear size of a clump of mass $M$ is
$r \propto \left( M / n m_p \right)^{1/3}$,
so the required resolution scales as $\Delta x \propto \GPE^{-1/3}$. A resolution of $\approx 4 \pc$ is needed for filling fractions to converge with a heating rate $\Gamma_{-25} = 0.85$ \citep{AvillezBreitschwerdt:2004,HillJoung:2012,KimOstriker:2017a}, so we expect that a resolution of $\approx 1.5 \pc$ is necessary for the highest heating rate we employ, $\Gamma_{-25} = 12.3$. We therefore include a $\Delta x_0 = 1 \pc$ {\tt pe1229} model.

\section{Hydrostatic equilibrium and pressure} \label{sec:hydrostatic}

In hydrostatic equilibrium, the total vertical pressure at height $z$ is balanced by the weight,
\begin{equation} \label{eq:weight}
W(z) = \int_z^\infty g(z') \rho(z') dz'.
\end{equation}
(The integral goes from $z$ to $-\infty$ for $z < 0$.) The vertical distribution of density $\rho(z)$ thus determines the midplane weight. The weight is not fixed for a given surface mass density but instead responds to the feedback processes in the disk. The calculation of the weight is not entirely straightforward. Isolated, high-altitude warm and cold clouds which are falling or rising in a hot halo as part of the a Galactic fountain will contribute significantly to the integral in equation~(\ref{eq:weight}). However, these clouds are moving supersonically, and thus cannot exert pressure forces beyond their bounding shocks and rarefaction waves, so they do not apply any force on gas in the midplane (E. Ostriker, private communication). Therefore, equation~(\ref{eq:weight}) overestimates the true weight felt by gas in the midplane, especially in cases in which there is a large mass of high-altitude warm or cold gas surrounded by hot gas and out of dynamical equilibrium.

With the physics included in our model, the total vertical pressure consists of the thermal pressure, the vertical turbulent pressure, and the sum of the magnetic pressure and tension \citep{Parker:1969,BoularesCox:1990},
\begin{eqnarray} \nonumber
\Ptotz &=& \Pth + \Pturb + P_B \\
&=& \langle nkT \rangle + \langle \rho \sigma_z^2 \rangle + \frac{\langle|\vec{B}|^2\rangle}{8\pi} - \frac{\langle B_z^2 \rangle}{4 \pi}.
\end{eqnarray}
Here $\sigma_z$ is the dispersion of the vertical speed $v_z$ calculated over some volume \citep[see][and discussion in Section~\ref{sec:turbulent_pressure} below]{JoungMac-Low:2009}.
In our models (which do not have shear or rotation), the net effect of the magnetic field on vertical support is small due to the arrangement of the horizontal and vertical components of the field \citep{GresselElstner:2008,HillJoung:2012}, so we neglect it.

\begin{figure*}[tb]
\includegraphics[width=\textwidth]{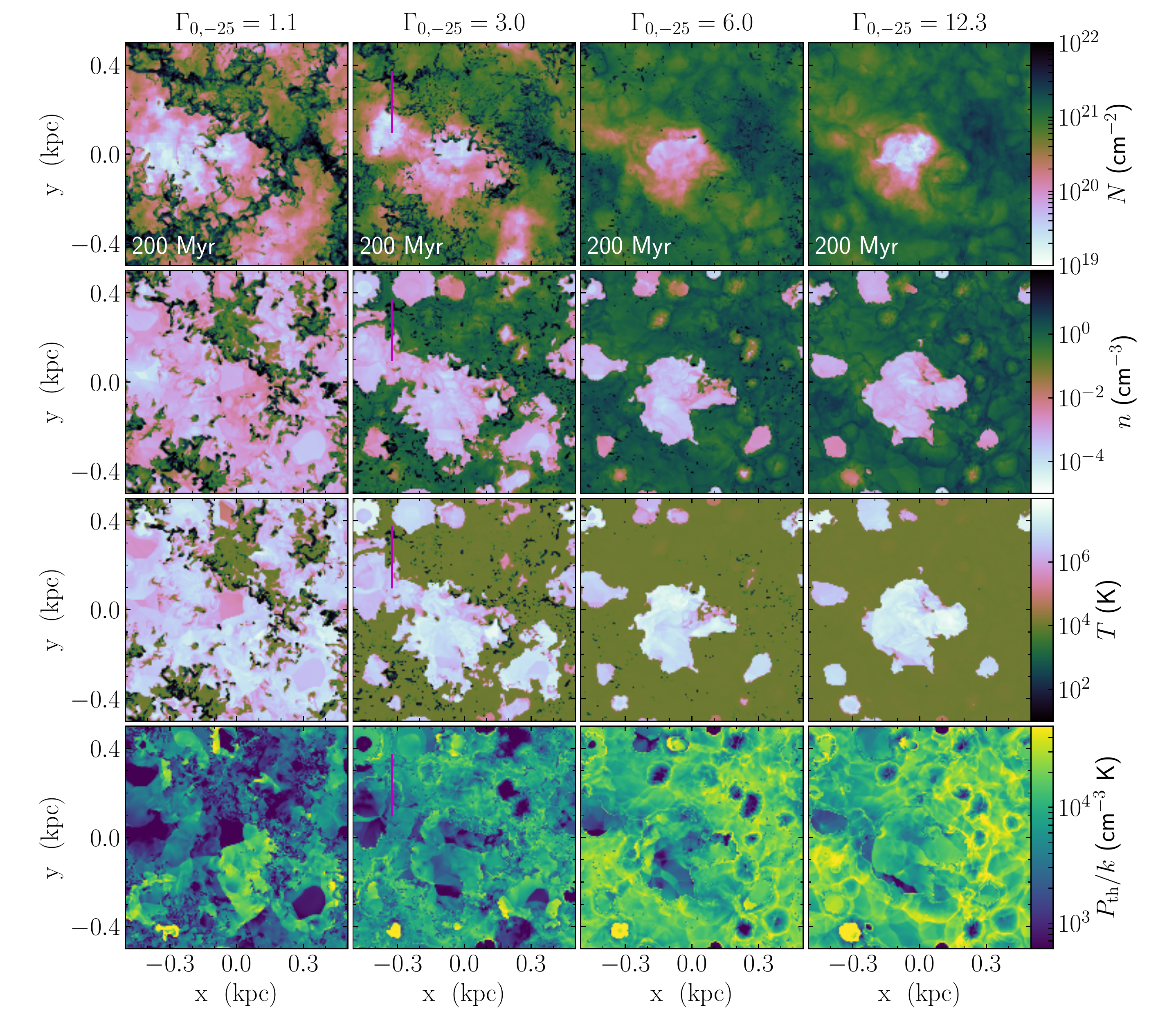}
\caption{Images of column density, number density, temperature, and thermal pressure (top to bottom) in the midplane ($z=0$) of $\Delta x_0 = 2 \pc$ simulations from runs with varying heating rates but all other inputs, including the supernova history, fixed (see Table~\ref{tbl:runs}). The fiducial model ($\Gamma_{0,-25} = 3.0$) is the second column. Each column is from a single snapshot with the heating rate $\Gamma_{0,-25}$ indicated. Magenta lines show the locations of the profiles in Figure~\ref{fig:T_pk_y_profile} below.}
\label{fig:images}
\end{figure*}

\begin{figure*}[tb]
\includegraphics[width=\textwidth]{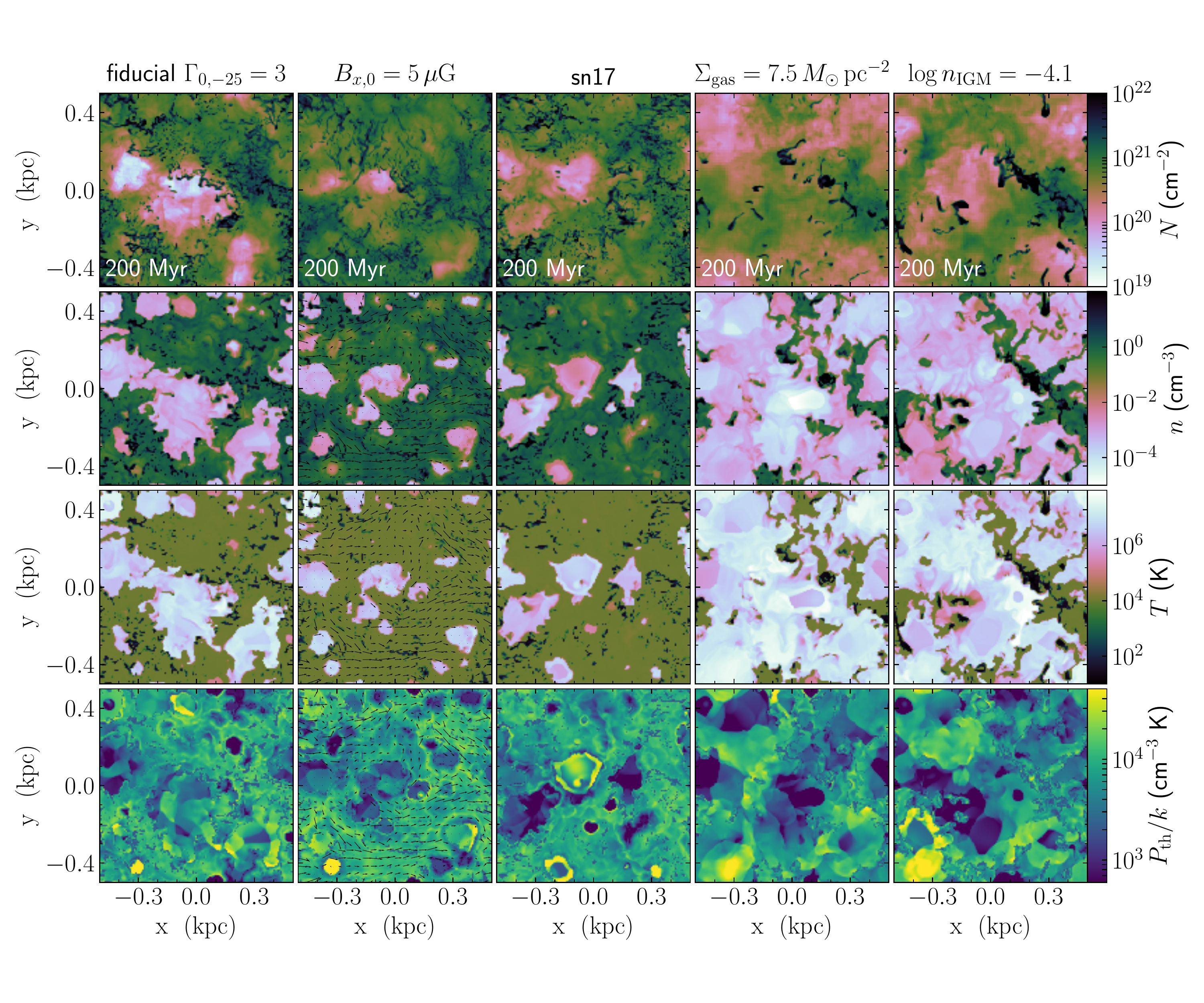}
\caption{Images of column density, density, temperature, and thermal pressure (top to bottom) in the midplane ($z=0 \pc$) of $\Delta x_0 = 4 \pc$ simulations from runs with $\Gamma_{0,-25} = 3$. The fiducial model, {\tt pe300}, is in the left column. The other columns are models with one parameter changed from the fiducial model: left to right, $B_{x,0} = 5 \uG$, half the core-collapse supernova rate, the surface mass density reduced from $13.2$ to $7.5 \mbox{ M}_\odot \pc^{-2}$, and the IGM density increased from $10^{-7.1}$ to $10^{-4.1} \cucm$. For the magnetized run, arrows show the planar magnetic field strength and direction.}
\label{fig:images_pe300}
\end{figure*}

\begin{figure*}[p]
\includegraphics[width=\textwidth]{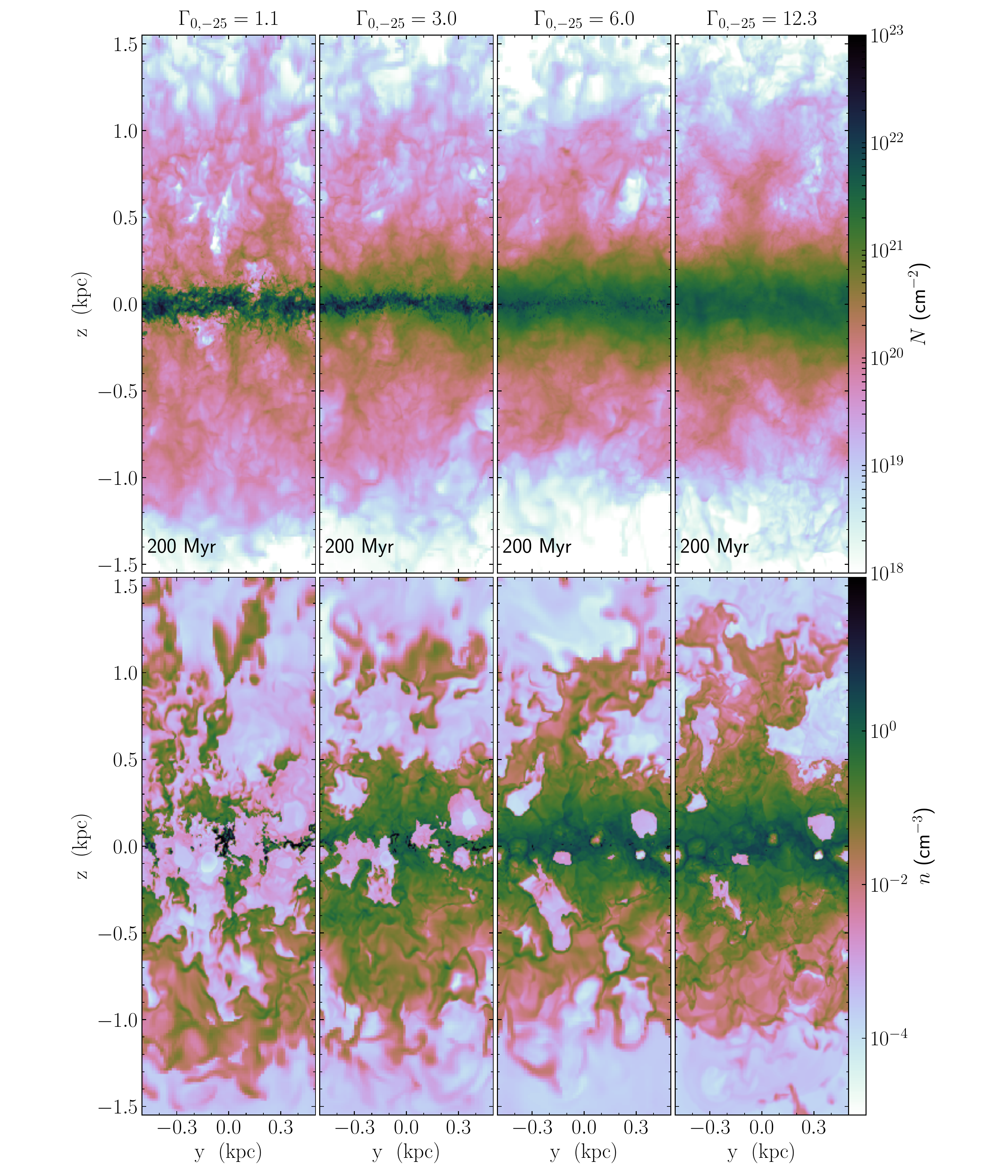}
\caption{Images of column density and number density in the $y-z$ plane ($x=+400 \pc$) of simulations from the runs in Figure~\ref{fig:images}.}
\label{fig:images_x}
\end{figure*}

\begin{figure*}[p]
\includegraphics[width=\textwidth]{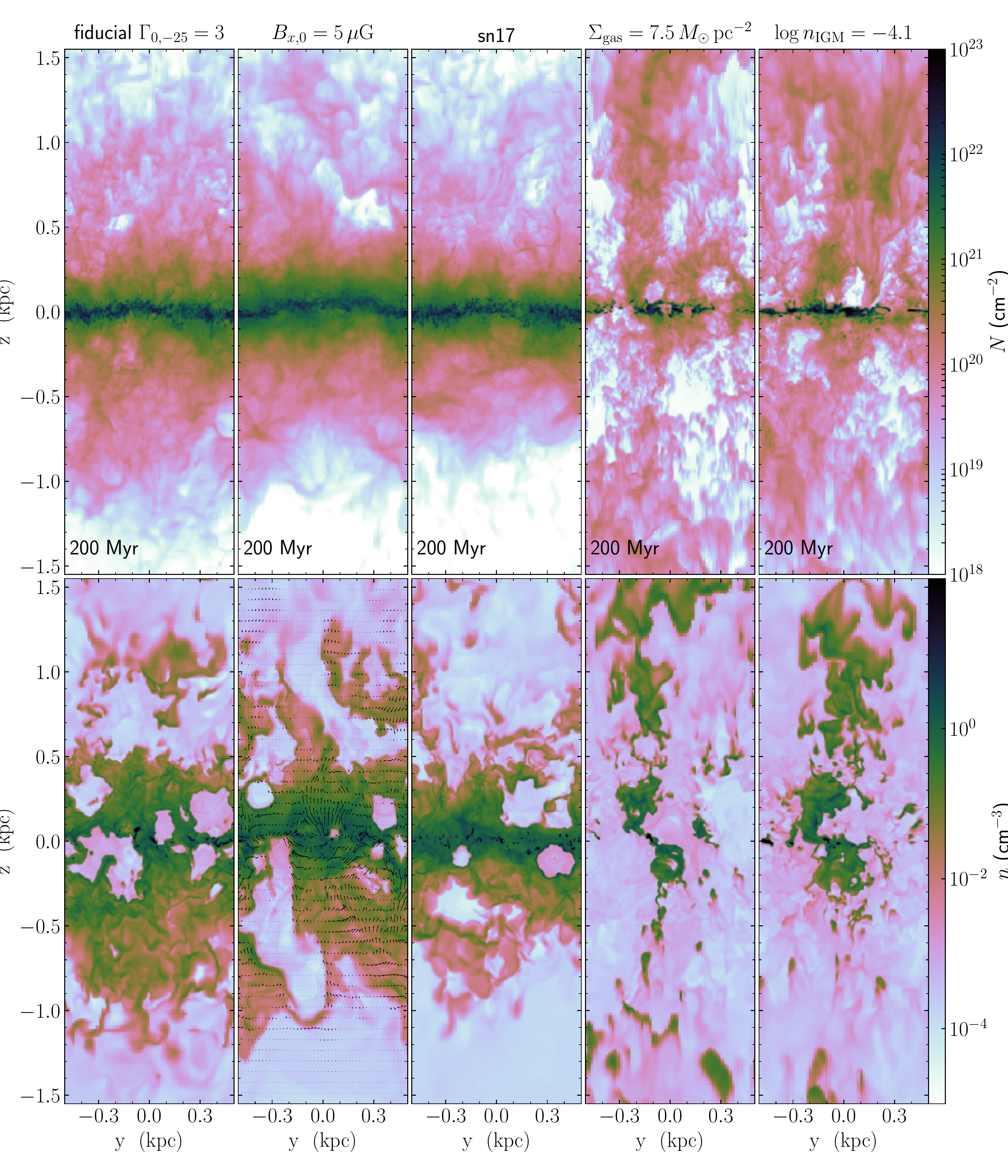}
\caption{Images of column density and density in the $y-z$ plane ($x=+400 \pc$) of simulations from runs in Figure~\ref{fig:images_pe300}, with magnetic field vectors shown in the density image for the magnetized model.}
\label{fig:images_pe300_x}
\end{figure*}

\section{Results} \label{sec:results}

\subsection{Description of models}

We begin with a qualitative description of the appearance of the runs based on column density images and slices of density, temperature, and thermal pressure in the midplane and vertical slices in Figures~\ref{fig:images}--\ref{fig:images_pe300_x}. Figure~\ref{fig:images} shows our fiducial series of runs, with all inputs equal except the heating rate, which varies over $1.1 \le \Gamma_{0,-25} \le 12.3$. In most of our stratified-disk models, a three-phase ISM is present. At low heating rates, the volume is dominated by the HIM with relatively-small pockets of WNM. There are long, filamentary regions of CNM. At high heating rates, the volume is dominated by the WNM; the HIM is confined to relatively-small supernova-driven bubbles and remnants. The CNM is confined to increasingly-small clouds and clumps with increasing heating rates. At the highest heating rate, very little CNM survives at all; the cold gas accounts for fractions $\lesssim 2 \times 10^{-3}$ of the mass and $\lesssim 10^{-5}$ of the volume in the midplane at $200 \Myr$ in the {\tt pe1229 2pc} and {\tt pe1229 1pc} models (see Table~\ref{tbl:outputs}).

\begin{figure}
\includegraphics[width=0.5\textwidth]{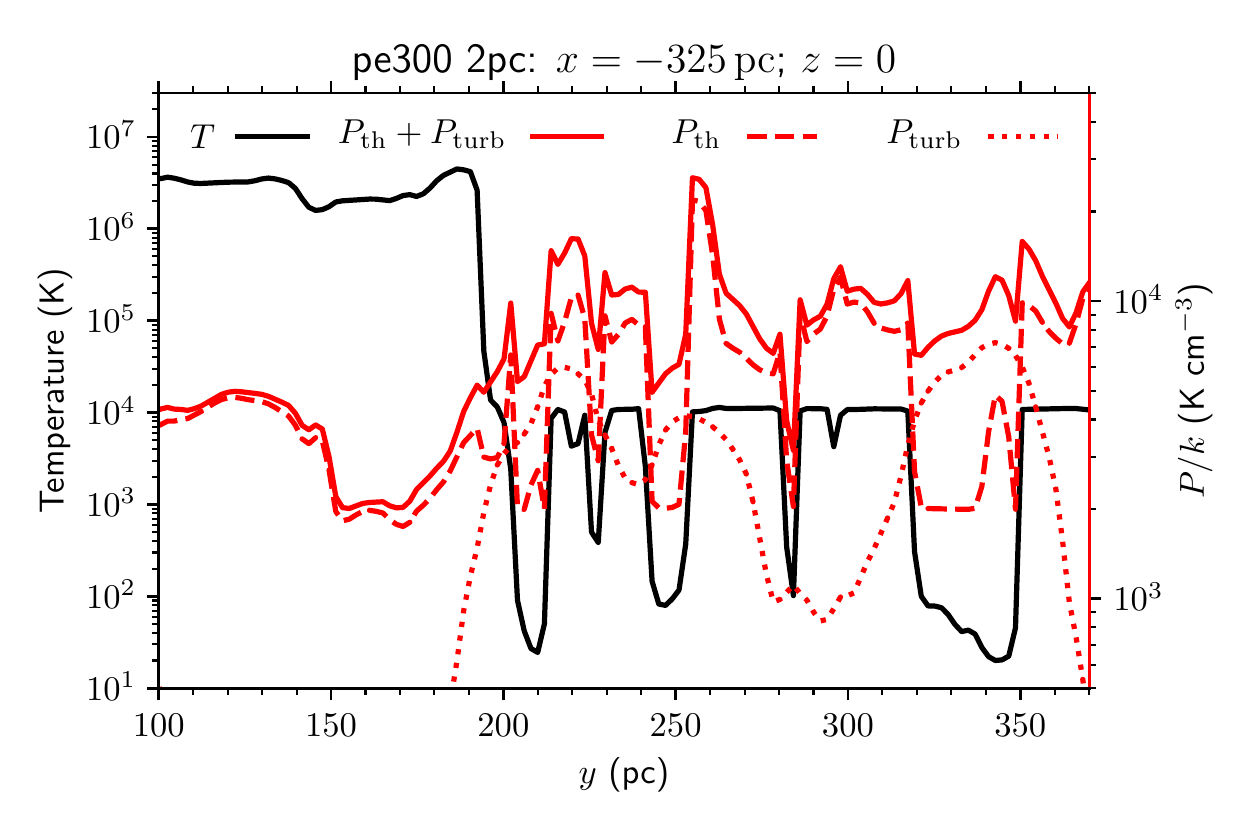}
\caption{Temperature (black; left axis) and thermal, turbulent, and total pressure (red; right axis) as a function of position along a line at $x=-325 \pc$, $z=0$ in the {\tt pe300 2pc} model (shown as magenta lines in Figure~\ref{fig:images}). These profiles show the interfaces between HIM, WNM, and CNM regions. The turbulent pressure is calculated in $15 \pc$ radius spheres around each coordinate, so turbulent pressures of nearby points are not independent.}
\label{fig:T_pk_y_profile}
\end{figure}

There is considerably less contrast in thermal pressure than in density and temperature, indicating that the ISM is in very approximate thermal pressure equilibrium. (Note the narrower range of the \Pth\ color scale in the images.) The midplane thermal pressure is higher at higher heating rates. The cold clouds are typically at lower thermal pressure, by a factor of $\approx 4$ in the {\tt pe300 2pc} model, than their immediate surroundings, which are typically warm. This is evident in Figure~\ref{fig:T_pk_y_profile}, a profile through a slice of the {\tt pe300 2pc} model which includes CNM, WNM, and HIM regions.

In Figure~\ref{fig:images_pe300}, we show five models, all with $\Gamma_{0,-25} = 3$ but with other inputs changed. First, we include a magnetic field. The WNM fills more of the volume and the HIM less of the volume in the magnetized simulation. Second, we reduce the supernova rate by a factor of two ({\tt sn17} models). This results in less energy injection in shocks, so the WNM again fills more of the volume than in the fiducial run. Third, we reduce the surface mass density by a factor of $1.8$ ({\tt smd75}). Because we keep the supernova rate and heating rate the same, this puts the effective star formation rate in the modeled galaxy well above the Kennicutt-Schmidt relation. This lower surface mass density run has a weight that is about $30\%$ {\em higher} than the fiducial run, and the volume is dominated by HIM, similar to the low-heating-rate $\Gamma_{0,-25} = 1.09$ run with the fiducial surface mass density density. Lastly, we increase the density in the IGM ({\tt nigm-4}). In this model, the HIM again dominates the volume.

\begin{figure}
\includegraphics[width=0.5\textwidth]{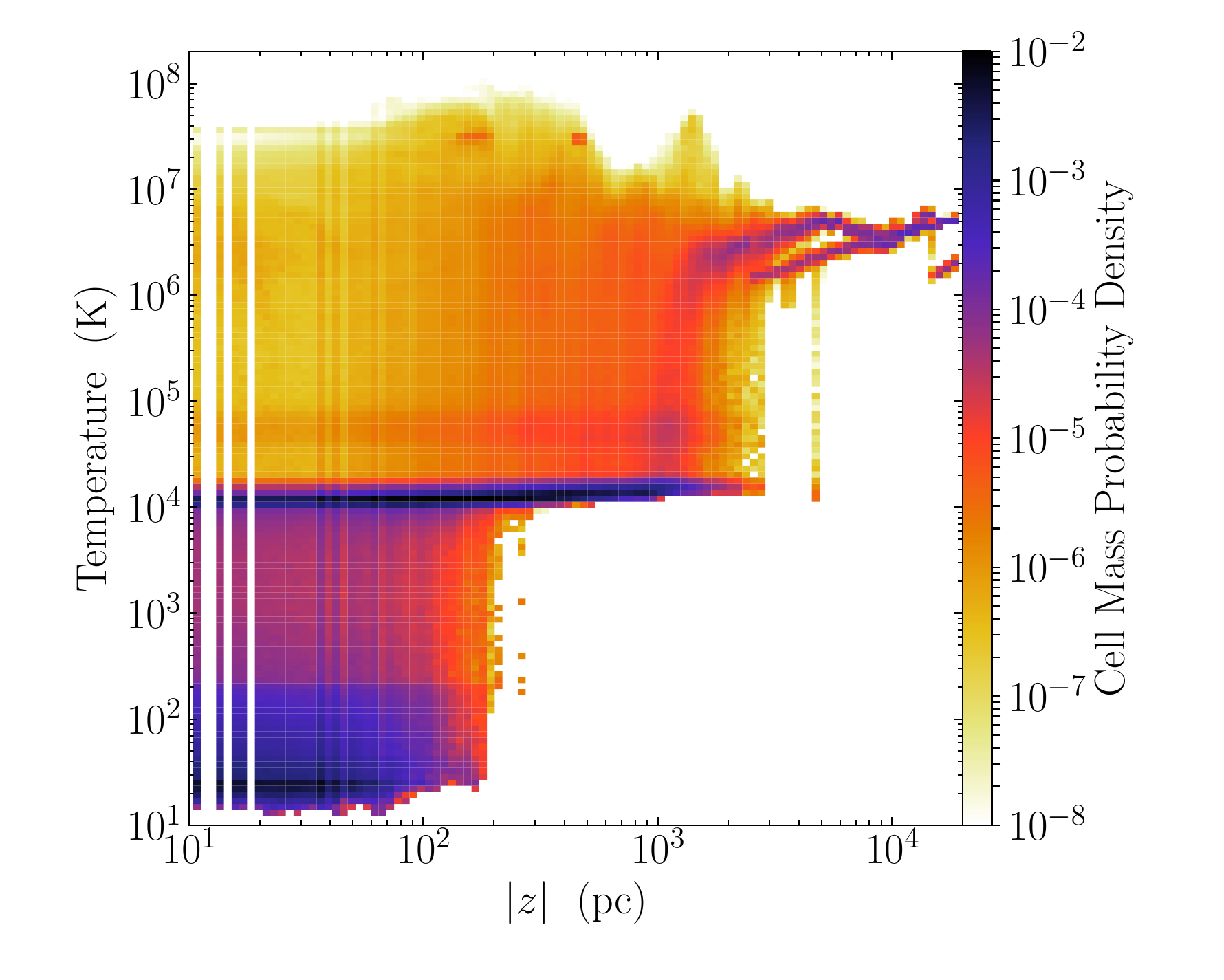}
\caption{Phase diagram of temperature as a function of $|z|$ (above and below the plane) in the fiducial {\tt pe300 2pc} model in the $t=200 \Myr$ snapshot. Bins are calculated with constant intervals of $\log |z|$, leading to bands with no data at small $|z|$.}
\label{fig:phase_T_z}
\end{figure}

\begin{figure*}[tb]
\plotone{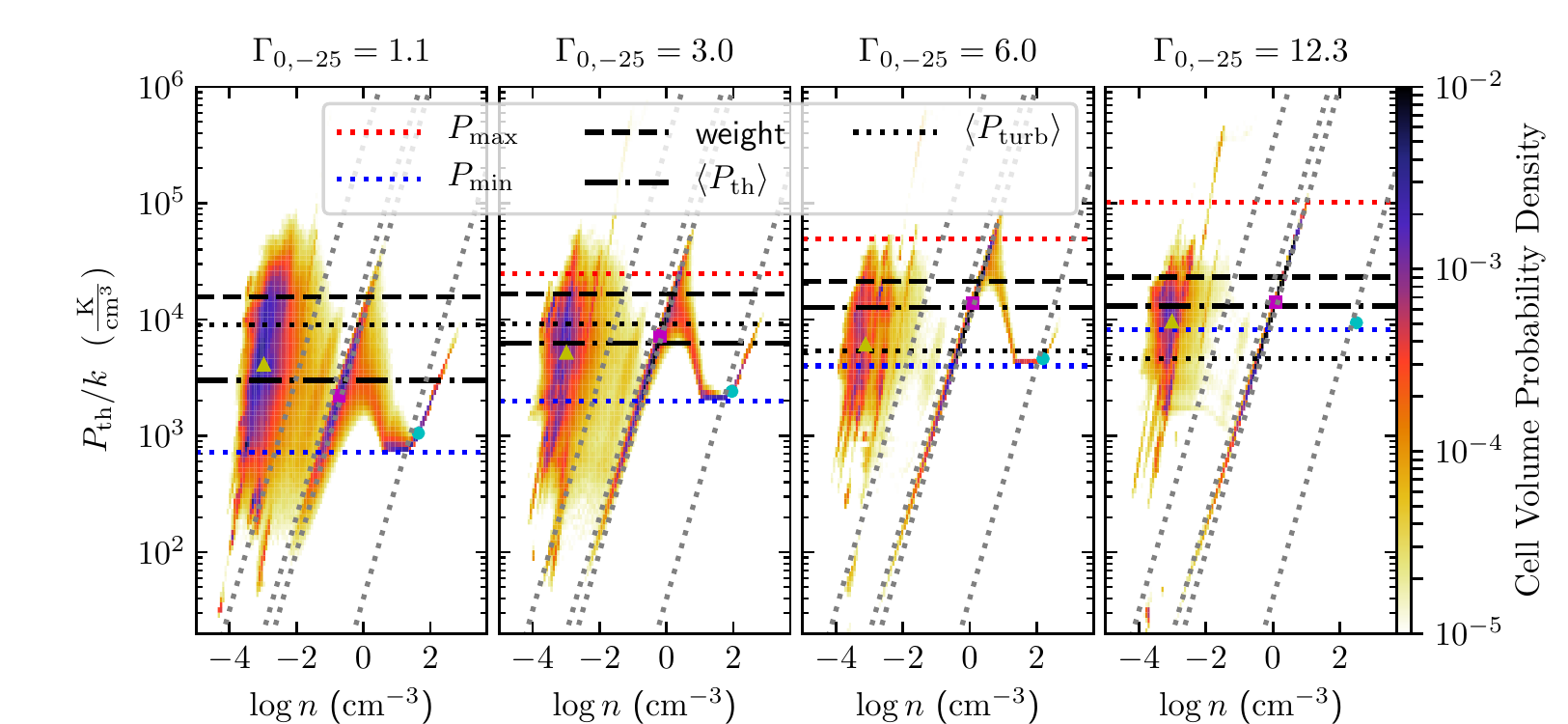}
\includegraphics[width=\textwidth]{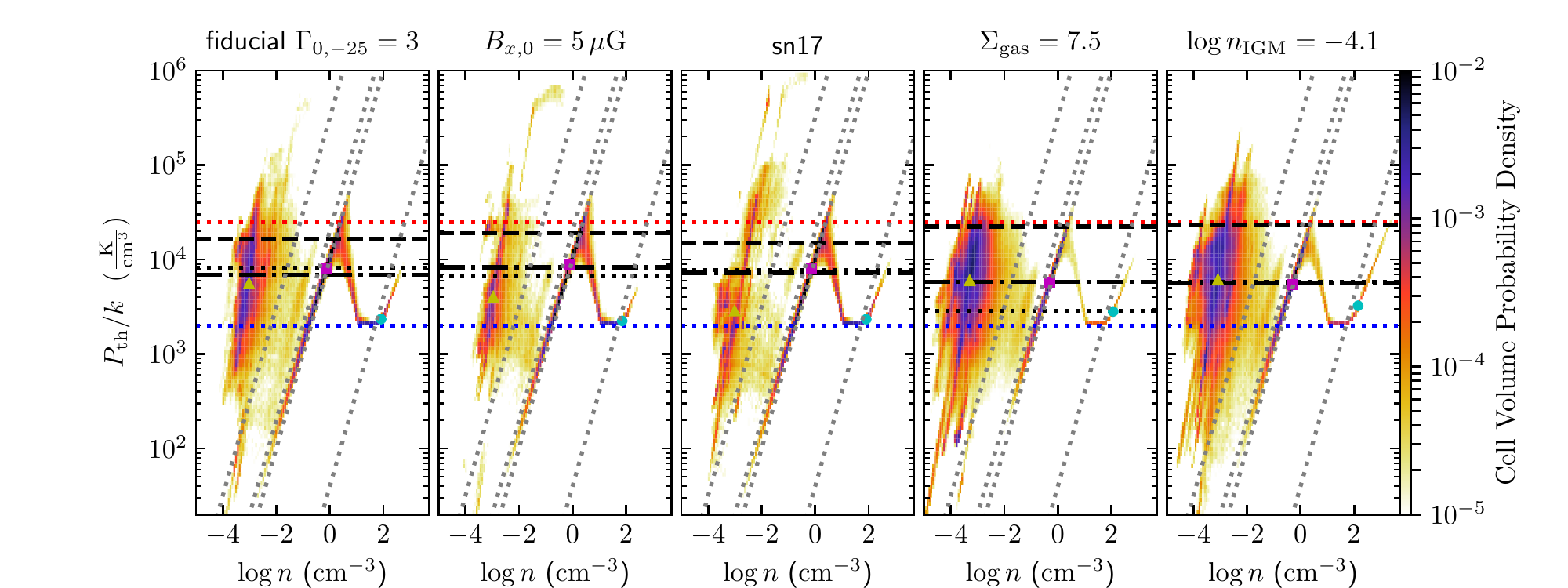}
\caption{Phase diagrams of thermal pressure as a function of density in the midplane ($|z| < h_z/3$) from single snapshots of the simulations shown in Figures~\ref{fig:images} and \ref{fig:images_x} (top) and Figures~\ref{fig:images_pe300} and \ref{fig:images_pe300_x} (bottom) at $t=200 \Myr$. Horizontal dashed black lines show the midplane hydrostatic pressure (the sum of the weight of the gas both above and below the midplane). Yellow triangles, magenta squares, and cyan circles show the median thermal pressure in each phase(hot, warm, and cold, respectively); dot-dashed black lines show the overall median thermal pressure; and dotted black lines show the median turbulent pressure (see Section~\ref{sec:turbulent_pressure}). Diagonal gray dotted lines denote the thermally unstable temperature ranges for our cooling curve. Red and blue dotted horizontal lines show \Pmax\ and \Pmin, respectively.}
\label{fig:phase}
\end{figure*}

How are the phases distributed vertically? We show vertical images in Figures~\ref{fig:images_x} and \ref{fig:images_pe300_x} and a phase diagram of the temperature as a function of height $|z|$ in Figure~\ref{fig:phase_T_z}. Near the midplane, all three phases coexist. The CNM is confined to $|z| \lesssim 200 \pc$; there is a mix of warm and hot gas in the range $200 \pc \lesssim |z| \lesssim 3 \kpc$ with gas existing at a wide range of temperatures. At larger heights, the gas is isothermal. (There are two narrow, isothermal curves evident in Figure~\ref{fig:phase_T_z}; one is at $z \gtrsim 3 \kpc$ and the other at $z \lesssim -3 \kpc$.) In this portion of the simulation domain, a single shock propagates upwards, filling the simulation domain horizontally. This one-dimensional behavior is an artificial consequence of the periodic horizontal boundary conditions and our tall aspect ratio \citep{WaltersCox:2001,GentShukurov:2013a}.

We characterize the vertical distribution of the gas with the mass-weighted rms altitude of the gas \citep{PiontekOstriker:2007},
\begin{equation}
h_z = \left(\frac{\sum_i z_i^2 \rho_i}{\sum_i \rho_i} \right)^{1/2}.
\end{equation}
For each simulation, we calculate statistics for $|z| < h_z / 3$ in order to confine our analysis to the region where the density and pressure are $\gtrsim 70\%$ of their midplane values. We show these statistics in Figures~\ref{fig:phase}--\ref{fig:pth_pturb} as well as Table~\ref{tbl:outputs}. Although our $4 \pc$ resolution models are converged in volume filling fractions and midplane pressure, the scale heights are not converged; $2 \pc$ resolution models typically have scale heights $\approx 55\%$ of those in the $4 \pc$ models.

\subsection{Relationship between pressure and porosity}

\begin{figure*}[tb]
\plotone{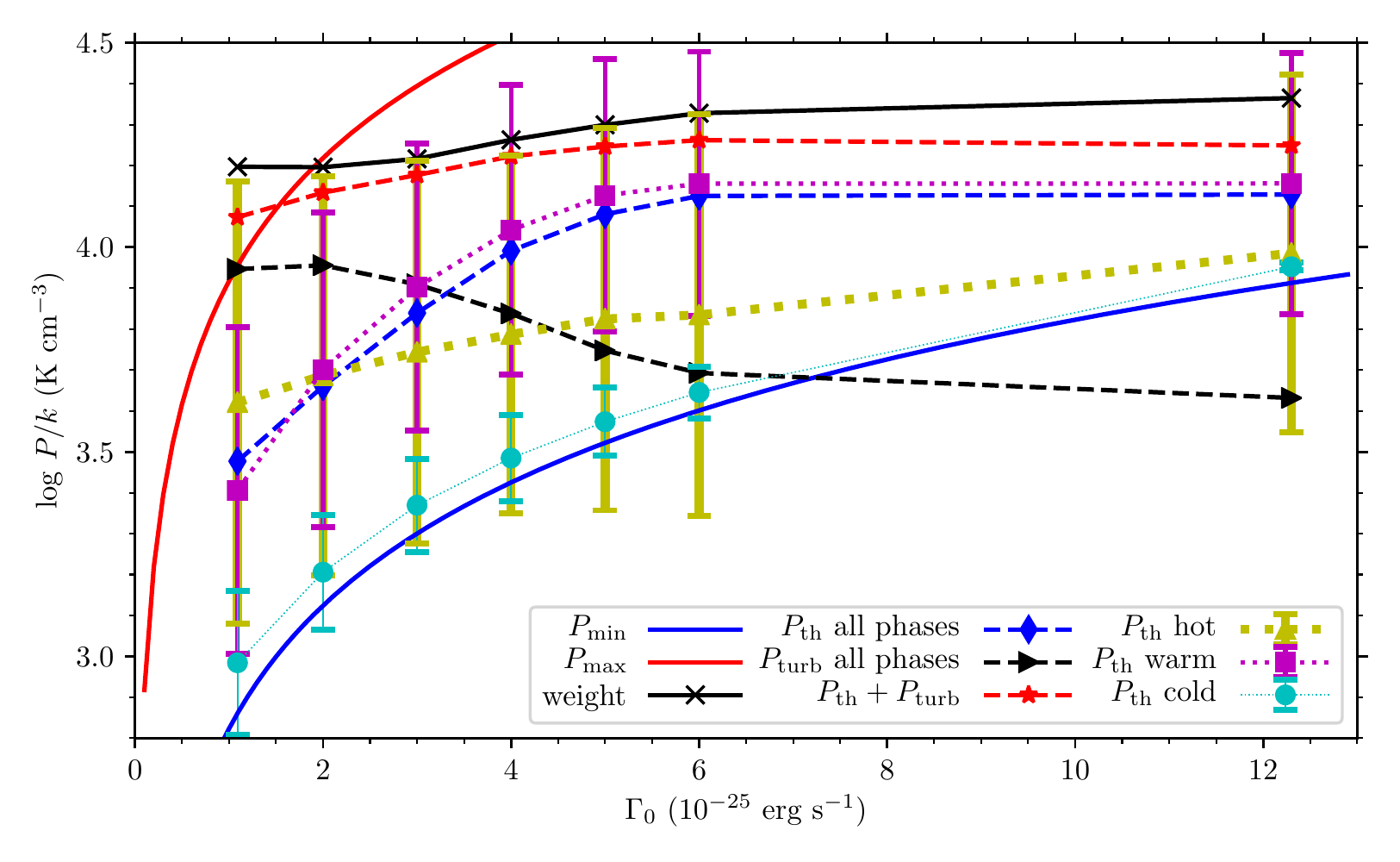}
\caption{Pressure as a function of $\Gamma_0$ at $200 \Myr$. The median thermal pressure within each phase is shown, as is the median overall thermal pressure, all calculated in the midplane ($|z| < h_z/3$). Error bars denote the standard deviation of $\log \Pth$ in the cold, warm, and hot phases; the standard error of the mean of each quantity is smaller than the symbol size. The median turbulent pressure (Section~\ref{sec:turbulent_pressure}) and the midplane weight are also shown. We show the fiducial models with $\Delta x_0 = 2 \pc$; input parameters in all models are identical except for $\Gamma_0$ (see Table~\ref{tbl:runs}). Red and blue lines show \Pmax\ and \Pmin, respectively (equation~\ref{eq:Pminmax}).}
\label{fig:p_gamma}
\end{figure*}

We show thermal pressure-density phase diagrams for all the models from Figures~\ref{fig:images}--\ref{fig:images_pe300_x} in Figure~\ref{fig:phase}. In all cases, a classical two-phase structure with a hot component in rough thermal pressure equilibrium is evident.
Less than $10^{-3}$ of the WNM has $\Pth > \Pmax$. In all models, the median is in the stable two-phase pressure regime, although WNM is found at pressures $\sim 1$~decade below \Pmin. The CNM is concentrated along the thermal equilibrium curve at or near $\Pth \approx \Pmin$. In the $\Gamma_{0,-25} = 12.3$ models, the mass and volume in the cold phase are essentially negligible, but what CNM there is is still found at $\Pth \approx \Pmin$. The thermally unstable cool and transition-temperature regimes are populated but fill considerably less volume than the thermally-stable branches or the HIM. The WNM pressure increases steadily as the heating rate increases.In contrast, the pressure of the HIM increases only slowly as the heating rate increases. Therefore, the HIM thermal pressure is higher than the median WNM thermal pressure in the models with low heating rates and lower than the median WNM thermal pressure in models with higher heating rates. We show this in a plot of the thermal pressure as a function of heating rate in Figure~\ref{fig:p_gamma}.

In most cases, the thermal pressure in each phase increases with $\GPE$ and, equivalently, \Pmin\ and \Pmax. With a Pearson correlation test, we can confidently reject the null hypothesis of a nonlinear correlation between the thermal pressure in each phase and $\Gamma_0$; this allows a linear correlation. The warm gas thermal pressure reaches a maximum of $\Pth/k \approx 1.4 \times 10^4 \K \cucm$ at $\Gamma_{-25} \approx 6$; at the highest heating rate, the thermal pressure is approximately the same. At these high heating rates, the thermal pressure of the warm gas is within $\approx 0.25$~dex of the weight of the ISM. The slope of the thermal pressure-heating rate relationship for the hot gas is shallow: the thermal pressure of the hot gas is higher than all the other phases at low $\GPE$ but lower than all but the cold phase at $\Gamma_{-25} \gtrsim 5$. At all heating rates where the CNM mass is significant, the median CNM thermal pressure is lower than that in any of the other phases.

In Figure~\ref{fig:p_gamma}, we also show the standard deviation of $\log \Pth$ in each phase as error bars. The scatter in the HIM thermal pressure is largest and that in the CNM thermal pressure is smallest, as is also evident from Figure~\ref{fig:phase}. The HIM thermal pressure range typically includes most of the pressure range in both the CNM and WNM. Moreover, the separation in thermal pressure between the CNM and WNM is substantial: in most cases, there is little overlap between $\langle \log \Pth \rangle \pm \sigma_{\log P\mathrm{th}}$ for the CNM and WNM. At the highest heating rate, the median thermal pressure in the CNM is within the range of the WNM thermal pressure, but this is unimportant for the ISM dynamics because the CNM accounts for nearly zero mass and volume in this model. In the high-porosity $\Gamma_{0,-25} = 1.1$ model, the median thermal pressure lies between the CNM and HIM thermal pressures and there is considerable overlap between the CNM and WNM thermal pressures.

\begin{figure}[tb]
\plotone{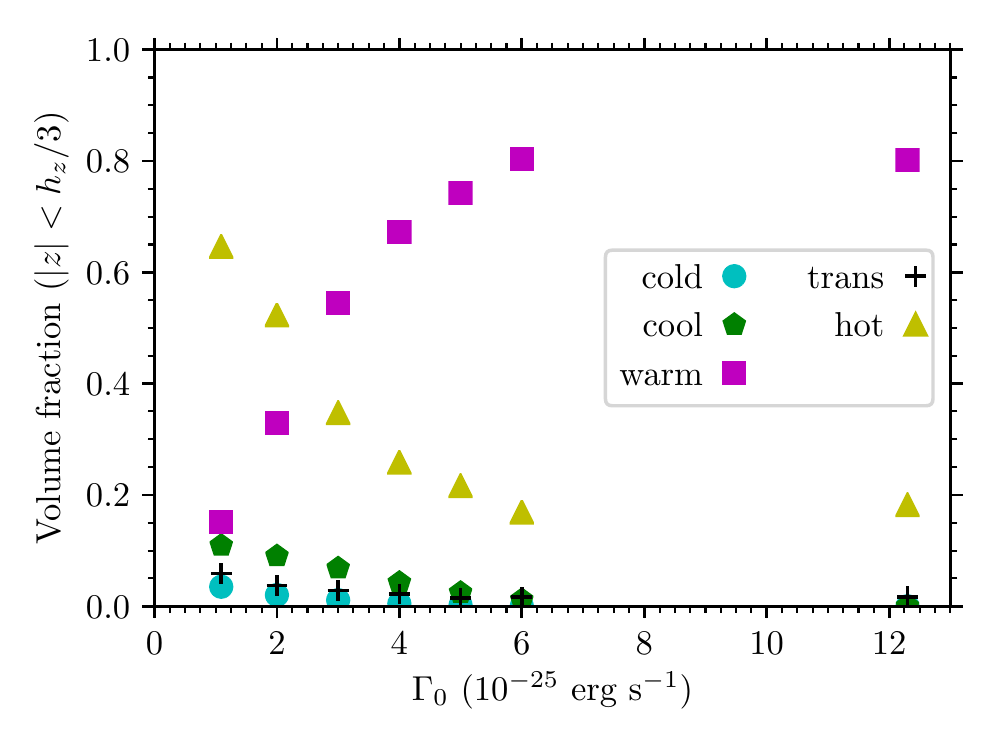}
\plotone{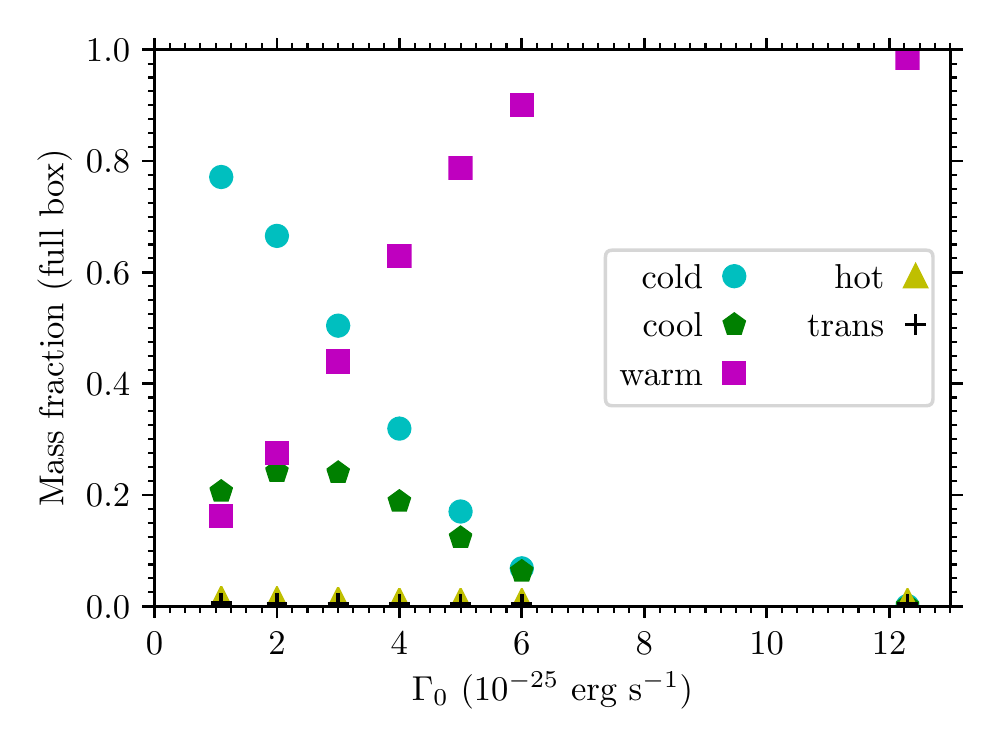}
\caption{Midplane ($|z| < h_z/3$) volume filling fraction and full-box mass fraction as a function of heating rates $\Gamma_0$ for each phase for the fiducial runs (as in Fig.~\ref{fig:p_gamma}).}
\label{fig:stats}
\end{figure}

\begin{figure}[tb]
\plotone{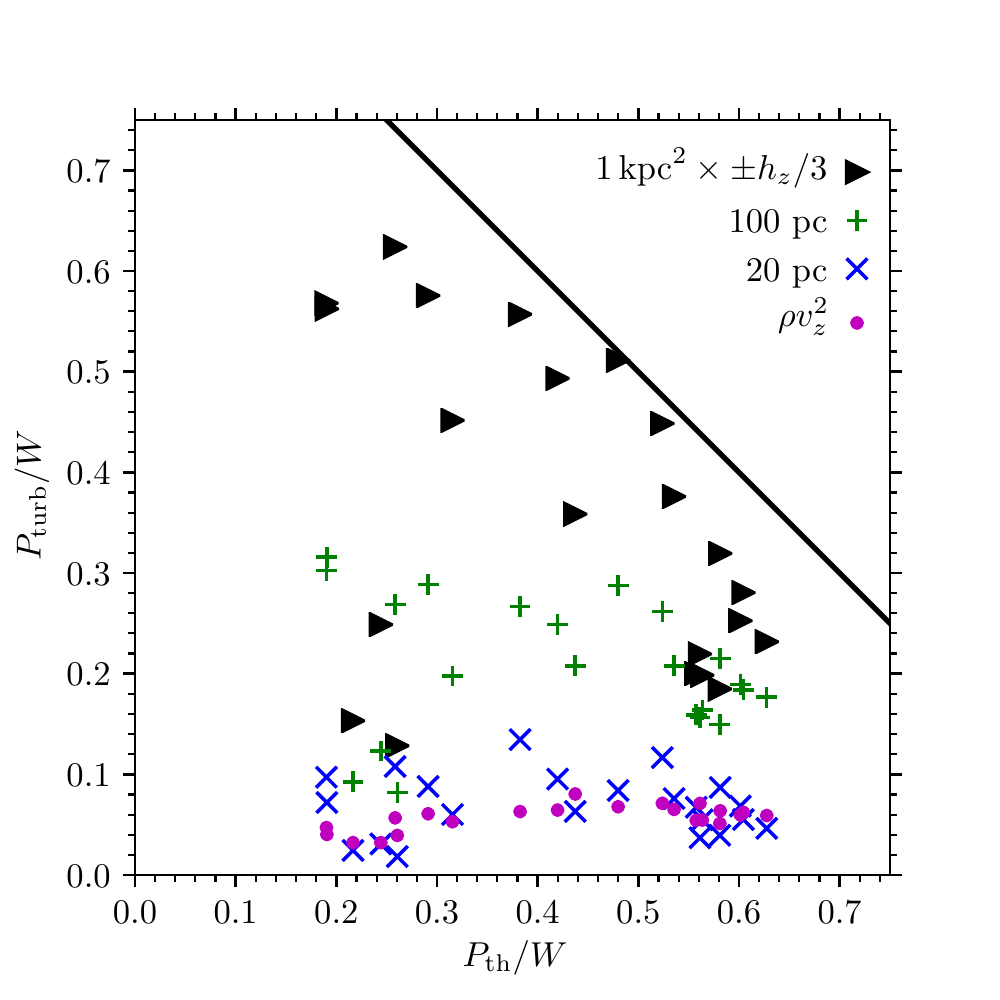}
\caption{Turbulent pressure as a function of thermal pressure, each normalized by the weight. The turbulent pressure is calculated over three different scales. The black triangles are $\langle \rho \rangle \sigma_z^2$ calculated over the $1\times1 \kpc^2$ box for $|z| < h_z/3$, as in Fig.~\ref{fig:p_gamma}. The $+$ and $\times$ signs are the median of $\langle \rho \rangle \sigma_z^2$ calculated for $\approx 2500$ spheres of the indicated radius centered in the midplane in the center-of-mass frame of the sphere. We also show $\langle \rho v_z^2 \rangle$ as magenta dots. If $\Pturb + \Pth = W$, the points would lie along the solid black $\Pth=\Pturb$ line.}
\label{fig:pth_pturb}
\end{figure}

In Figure~\ref{fig:stats}, we plot the volume filling factor and mass fraction in each of the five temperature ranges as a function of the heating rate. At low heating rates, the hot gas dominates by volume. The warm gas and hot gas volume filling factors are approximately equal at $2 \lesssim \Gamma_{0,-25} \lesssim 3$. At higher heating rates ($\Gamma_{0,-25} \gtrsim 5$), the warm gas fills $\approx 80\%$ and the hot gas $\approx 20\%$ of the volume. The mass is dominated by cold gas at low heating rates and warm gas at high heating rates. The unstable cool gas accounts for up to $\approx 20\%$ of the volume and mass but never more; most of this unstable gas is the marginally-stable gas. In all cases, $\lesssim 10\%$ of the volume of the ISM is cold gas, while $\lesssim 20\%$ of the mass is hot gas.

\begin{deluxetable*}{l DDDDDDDDD}
\tablecaption{Outputs
\label{tbl:outputs}}
\tabletypesize{\footnotesize}
\decimals
\tablehead{\colhead{Run Name} & \twocolhead{$h_z$} & \twocolhead{$f_{m,c}$} & \twocolhead{$f_{v,w}$} & \twocolhead{$q$} & \twocolhead{\nbarwc} & \twocolhead{$\log W/k$} & \twocolhead{$\log \Pturb/k$} & \twocolhead{$\log \Pth/k$} & \twocolhead{$\log \Ptot/k$} \\
\colhead{} & \twocolhead{(pc)} & \twocolhead{} & \twocolhead{} & \twocolhead{} & \twocolhead{($\mathrm{cm}^{-3}$)} & \multicolumn{8}{c}{$(\mathrm{pc} \cucm)$}}
\colnumbers
\startdata
\multicolumn{17}{c}{Fiducial} \\
\hline
{\tt pe109                   } & 131.8 & 0.77 & 0.15 & 1.04 & 4.8 & 4.20 & 3.95 & 3.48 & 4.07 \\ 
{\tt pe109 2pc               } &  69.0 & 0.78 & 0.13 & 1.15 & 8.3 & 4.20 & 3.95 & 3.48 & 4.08 \\ 
{\tt pe200                   } & 136.7 & 0.67 & 0.33 & 0.74 & 2.9 & 4.20 & 3.96 & 3.66 & 4.13 \\ 
{\tt pe200 2pc               } &  71.7 & 0.68 & 0.30 & 0.78 & 4.8 & 4.20 & 3.99 & 3.61 & 4.14 \\ 
{\tt pe300                   } & 139.8 & 0.50 & 0.55 & 0.43 & 1.8 & 4.22 & 3.91 & 3.84 & 4.18 \\ 
{\tt pe300 2pc               } &  76.7 & 0.53 & 0.52 & 0.45 & 2.9 & 4.22 & 3.96 & 3.80 & 4.19 \\ 
{\tt sn17 pe300                   } & 121.7 & 0.48 & 0.75 & 0.16 & 1.6 & 4.18 & 3.89 & 3.86 & 4.17 \\ 
{\tt pe400                   } & 147.3 & 0.32 & 0.67 & 0.30 & 1.5 & 4.26 & 3.84 & 3.99 & 4.22 \\ 
{\tt pe400 2pc               } &  84.7 & 0.36 & 0.64 & 0.33 & 2.1 & 4.26 & 3.91 & 3.98 & 4.25 \\ 
{\tt pe500                   } & 153.7 & 0.17 & 0.74 & 0.24 & 1.3 & 4.30 & 3.75 & 4.08 & 4.25 \\ 
{\tt pe500 2pc               } &  91.4 & 0.21 & 0.73 & 0.25 & 1.7 & 4.30 & 3.80 & 4.06 & 4.25 \\ 
{\tt pe600                   } & 156.9 & 0.07 & 0.80 & 0.18 & 1.2 & 4.33 & 3.69 & 4.13 & 4.26 \\ 
{\tt pe600 2pc               } &  96.8 & 0.10 & 0.79 & 0.20 & 1.5 & 4.33 & 3.73 & 4.11 & 4.26 \\ 
{\tt pe1229                  } & 166.1 & 0.00 & 0.80 & 0.20 & 1.0 & 4.36 & 3.63 & 4.13 & 4.25 \\ 
{\tt pe1229 2pc              } & 163.1 & 0.00 & 0.81 & 0.20 & 1.1 & 4.37 & 3.66 & 4.12 & 4.25 \\ 
{\tt pe1229 1pc              } & 106.4 & 0.00 & 0.81 & 0.20 & 1.3 & 4.37 & 3.67 & 4.11 & 4.25 \\ 
\hline
\multicolumn{17}{c}{Magnetized} \\
\hline
{\tt bx5 pe1229              } & 164.6 & 0.00 & 0.82 & 0.19 & 1.0 & 4.37 & 3.71 & 4.12 & 4.27 \\ 
{\tt bx5 pe300               } & 142.1 & 0.31 & 0.76 & 0.16 & 1.3 & 4.28 & 3.83 & 3.92 & 4.18 \\ 
\hline
\multicolumn{17}{c}{$n_\mathrm{IGM} = 10^{-4.1} \cucm$} \\
\hline
{\tt nigm-4 bx0 pe300        } & 201.2 & 0.61 & 0.17 & 1.51 & 4.9 & 4.37 & 3.76 & 3.75 & 4.06 \\ 
{\tt nigm-4 bx5 pe300        } & 165.1 & 0.44 & 0.57 & 0.43 & 1.5 & 4.29 & 3.94 & 3.79 & 4.17 \\ 
\hline
\multicolumn{17}{c}{$\Sigma_\mathrm{gas}$ $= 7.5 \, M_\odot \pc^{-2}$} \\
\hline
{\tt smd75 pe300                    } & 311.2 & 0.36 & 0.12 & 1.89 & 1.8 & 4.35 & 3.46 & 3.76 & 3.94 \\ 
{\tt smd75 pe600                    } & 320.6 & 0.23 & 0.24 & 1.35 & 1.0 & 4.37 & 3.56 & 3.71 & 3.94 \\ 
\enddata
\tablecomments{Output statistics for the runs listed in Table~\ref{tbl:runs} with $\Delta x_0 \le 4 \pc$ at $t=200 \Myr$. Quantities in columns 4--10 are calculated for $|z| \le h_z / 3$. Shown are the mass-weighted rms altitude (column 2), the cold gas mass fraction (3), the warm gas volume filling factor (4), the porosity (5), the mean density in $T < 2 \times 10^4 \K$ gas (6), the weight (7), the turbulent pressure calculated over the $1 \kpc^2 \times \pm h_z/3$ box (8), the thermal pressure (9), and the sum of the turbulent and thermal pressures (10).}
\end{deluxetable*}

\subsection{Turbulent pressure and vertical support} \label{sec:turbulent_pressure}

We estimate the turbulent component of the vertical pressure by calculating the dispersion $\sigma_z$ of $v_z$ in the center-of-mass frame in regions of different sizes \citep[see][]{JoungMac-Low:2009}. The estimated vertical turbulent pressure is then $\Pturb = \langle \rho \rangle \sigma_z^2$, where $\langle \rho \rangle$ is the mean density in each region. This procedure averages over all gaseous phases.

Turbulent pressure is a scale-dependent quantity. The ISM turbulent spectrum is observed to extend to at least kiloparsec scales in external galaxies \citep[e.g.][and references therein]{WillettElmegreen:2005}, although the largest scales are likely due to self-gravity of the stars and gas, effects which we do not include in our model. The processes we include produce a maximum scale of around $100 \pc$ \citep{JoungMac-Low:2006,AvillezBreitschwerdt:2007}.
Therefore, we also calculate the turbulent pressure in different volumes. We calculated $\sigma_z$ within spheres of radius $20 \pc$ and $100 \pc$, giving us $\langle \rho \rangle \sigma_z^2$ for $\approx 400$ spheres centered at different locations in the midplane ($z=0$). We show the median of these 400 turbulent pressure measurements for each model as the $+$ and $\times$ signs in Figure~\ref{fig:pth_pturb}. The turbulent pressure is larger in the larger boxes. We also include the median of $\rho v_z^2$ as an estimate of the kinetic pressure in Figure~\ref{fig:pth_pturb}. This value is small compared to turbulent pressure calculated from $\sigma_z$ and independent of thermal pressure, with $\langle \rho v_z^2 \rangle / W \approx 0.05$ in all models.

In Figure~\ref{fig:pth_pturb}, we plot the turbulent pressure as a function of thermal pressure normalized by the total weight in each simulation. The black right-facing triangles in Figures~\ref{fig:p_gamma} and \ref{fig:pth_pturb} use the $1 \kpc^2 \times \pm h_z/3$ box. The turbulent pressure is $4000 \K \cucm \lesssim \Pturb / k \lesssim 10^4 \K \cucm$. The turbulent pressure is larger at low heating rates than at high heating rates, although the relationship is not monotonic (Fig.~\ref{fig:p_gamma}). At the lowest heating rate, the turbulent pressure is greater than the thermal pressure in all phases, whereas at the highest heating rate, the turbulent pressure is lower than the thermal pressure in all phases.

The median turbulent and thermal pressures are inversely correlated.\footnote{We note that because we calculate \Pturb\ in boxes extending to $h_z/3$, the volume over which this pressure is calculated changes between models. In general, larger volumes lead to larger turbulent pressures (Fig.~\ref{fig:pth_pturb}). This effect does not explain the anticorrelation between thermal and turbulent pressures: at lower heating rates, $h_z$ is smaller, yet the calculated turbulent pressure is higher (Fig.~\ref{fig:stats}).} If the vertical support were provided fully by the turbulent and thermal pressure, we would have
\begin{equation} W = \Pth + \Pturb, \end{equation} 
which can be checked by comparing black $\times$ symbols and red stars in Figure~\ref{fig:p_gamma}, and by examining how closely the symbols approach
the solid line in Figure~\ref{fig:pth_pturb}. For the fiducial models, in which we vary only $\Gamma_0$, we find$(\Pth + \Pturb)/W > 0.85$, so these figures suggest that our estimates of the thermal and turbulent pressure capture most of the hydrostatic support of the ISM. Because we only estimate the turbulent pressure on a single (large) scale, it is reasonable to expect that we are not capturing all of the turbulent pressure. We may also overestimate the weight (Section~\ref{sec:hydrostatic}).

Figure~\ref{fig:pth_pturb} and Table~\ref{tbl:outputs} show that the turbulent and thermal pressure account for most of the hydrostatic support in most of the runs. There are three data points that are clear outliers, with $\Pturb/W < 0.3$ and $\Pth/W < 0.3$. Two of these models are the low surface mass density models ({\tt smd75 pe300} and {\tt smd75 pe600}). The third is the unmagnetized model with the increased IGM density ({\tt nigm-4 pe300}). These are the three models with the largest scale heights and the highest porosities (see the right two panels of Figure~\ref{fig:images_pe300_x}) and are qualitatively different from the other models presented in this paper: the cold and warm disk is essentially blown away, and much of the mass is concentrated in high-altitude clouds that do not contribute to the midplane pressure (see Section~\ref{sec:hydrostatic}). We conclude that in all of our models in which the ISM has a disk-like structure (as is observed in star-forming galaxies), our method for estimating the thermal and turbulent pressure captures $\gtrsim 85\%$ of the hydrostatic support and the ISM is in approximate hydrostatic equilibrium.

In Figure~\ref{fig:T_pk_y_profile}, we show a profile of the temperature and thermal and turbulent pressure along a ray through the midplane of our fiducial model. We chose the position of this profile to include several CNM regions embedded in WNM gas as well as the edge of a HIM bubble. To illustrate the turbulent pressure on small scales, we calculated the turbulent pressure in spheres of radius $15 \pc$, chosen to balance between minimizing the size scales over which we smooth the turbulent pressure and maximizing the turbulent energy we capture. The CNM and WNM thermal pressures here are each near the median values for this model (see Figure~\ref{fig:p_gamma}). The HIM thermal pressure varies by an order of magnitude around its median.

Across the CNM-WNM interface, the total pressure is generally more flat than either the turbulent or thermal pressure: the turbulent pressure and thermal pressure are generally inversely correlated across CNM regions at least as large as the $30 \pc$ diameter kernel we used to calculate \Pturb. This effect is most clear in the CNM region at $320 \pc \lesssim y \lesssim 350 \pc$. This is a local equivalent of the global anti-correlation between \Pth\ and \Pturb\ described above: the thermal pressure is lower in the CNM than the WNM but the turbulent pressure appears to adjust to maintain rough {\em total} pressure equilibrium between the phases.

Because turbulent motions are spatially correlated, \citet{WolfireMcKee:2003} argued that it is the thermal pressure that determines the phase structure on scales $\lesssim 200 \pc$ in the WNM. However, they assumed a much longer correlation length than was found in subsequent simulations \citep{AvillezBreitschwerdt:2004,JoungMac-Low:2006}. Our analysis here indicates that the turbulent pressure does act on the scale of individual CNM-WNM interfaces. We caution that our CNM clouds are only marginally resolved and the CNM-WNM interface, which forms on \citet{Field:1965jb} length scales, is entirely unresolved \citep{BegelmanMcKee:1990,InoueInutsuka:2006,StoneZweibel:2009}.

\section{Discussion} \label{sec:discussion}

In this paper, we have presented models of a supernova-driven turbulent ISM in a vertical column with a range of input heating rates. Changing the heating rate changes \Pmin\ and \Pmax, the limits of the pressure range in which the two-phase ISM is stable. We find that the thermal pressure in the cold and warm gas is also set by the heating rate, increasing directly proportionally to \Pmin\ (Fig.~\ref{fig:p_gamma}). The weight, HIM thermal pressure, and turbulent pressure vary more weakly with the input heating rate. The WNM thermal pressure approaches an asymptote $\approx 0.25$~dex below the weight when $\Gamma_{0,-25} \gtrsim 5$.

The warm gas dominates at high heating rates, where $P_\mathrm{th,warm} > P_\mathrm{th,hot}$ and $\Pturb < \Pth$. The hot gas dominates at low heating rates, where $P_\mathrm{th,warm} \lesssim P_\mathrm{th,hot}$ and $\Pturb > \Pth$. The warm and hot gas filling fractions are approximately equal when the total thermal and turbulent pressures and the warm and hot gas thermal pressures are also approximately equal.

In all of our models, {\em the CNM and WNM are each found within the two-phase thermal pressure range even though we do not include self-regulation and a variable star formation rate}. Therefore, the gas physics must act to keep the pressure in the two-phase regime. Next, we explore the underlying physical processes.

\subsection{Stability of the three-phase ISM} \label{sec:stability}

Why does the ISM tend towards $\Pmin < \Pth < \Pmax$? \citet{WolfireMcKee:2003} pointed out that analysis of a pressure-density phase diagram defines five regimes in the CNM-WNM equilibrium. These are delineated by the mean density in atomic gas\footnote{$T < 2 \times 10^4 \K$ for our cooling curve.} \nbarwc\ relative to, in increasing order of density, $n_\mathrm{WNM}(\Pmin)$, $n_\mathrm{WNM}(\Pmax)$, $n_\mathrm{CNM}(\Pmin)$, and $n_\mathrm{CNM}(\Pmax)$; the numerical values of these densities are set by the cooling curve and the heating rate. In these regimes, in order of increasing density, 1) only WNM gas can exist stably, 2) the ISM is dominated by the WNM but CNM gas can exist at the same pressure, 3) both the CNM and WNM {\em must} exist because \nbarwc\ is thermally unstable, 3) the ISM is dominated by the CNM but WNM gas can exist at the same pressure, and 5) only the CNM can exist stably. We now apply the \citet{WolfireMcKee:2003} argument to our simulations.

\begin{figure}
\includegraphics[width=0.5\textwidth]{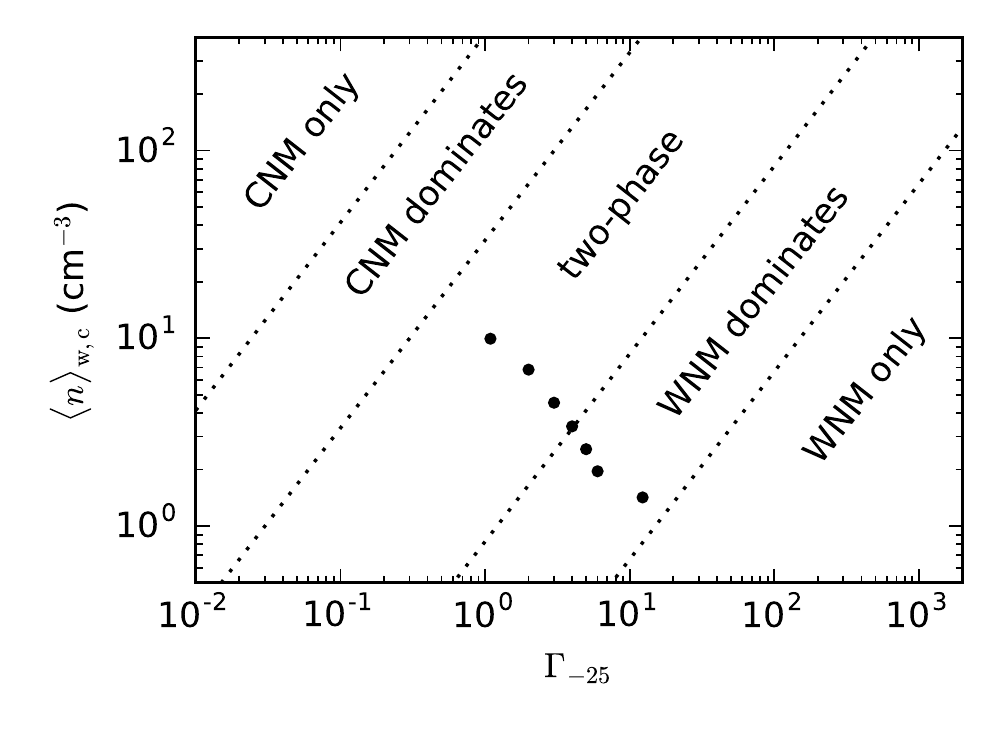}
\caption{Mean density as a function of heating rate. The five regimes described by \citet{WolfireMcKee:2003} (for the cooling curve used in this paper) are labeled. Points show \nbarwc\ from each of our base models.}
\label{fig:meann}
\end{figure}

We calculate \nbarwc\ for $|z| < h_z/3$, yielding the results in Table~\ref{tbl:outputs}. With fixed $\Sigma_\mathrm{gas}$, \nbarwc\ changes with increasing $\GPE$ due to the increasing scale height of the gas and the decreasing HIM volume filling fraction.
We show these five regimes along with the mean densities from our models as a function of $\GPE$ in Figure~\ref{fig:meann}. In all cases, the mean density is in the regime in which the thermal pressure allows the CNM and WNM to exist. In $\Gamma_{0,-25} \gtrsim 4$ models, $\nbarwc \lesssim n_\mathrm{WNM}(\Pmax)$, so the mean density lies along the stable WNM branch and we expect the WNM to dominate. Indeed, these are the models in which the CNM accounts for a minority of the mass.

\begin{figure}
\includegraphics[width=0.5\textwidth]{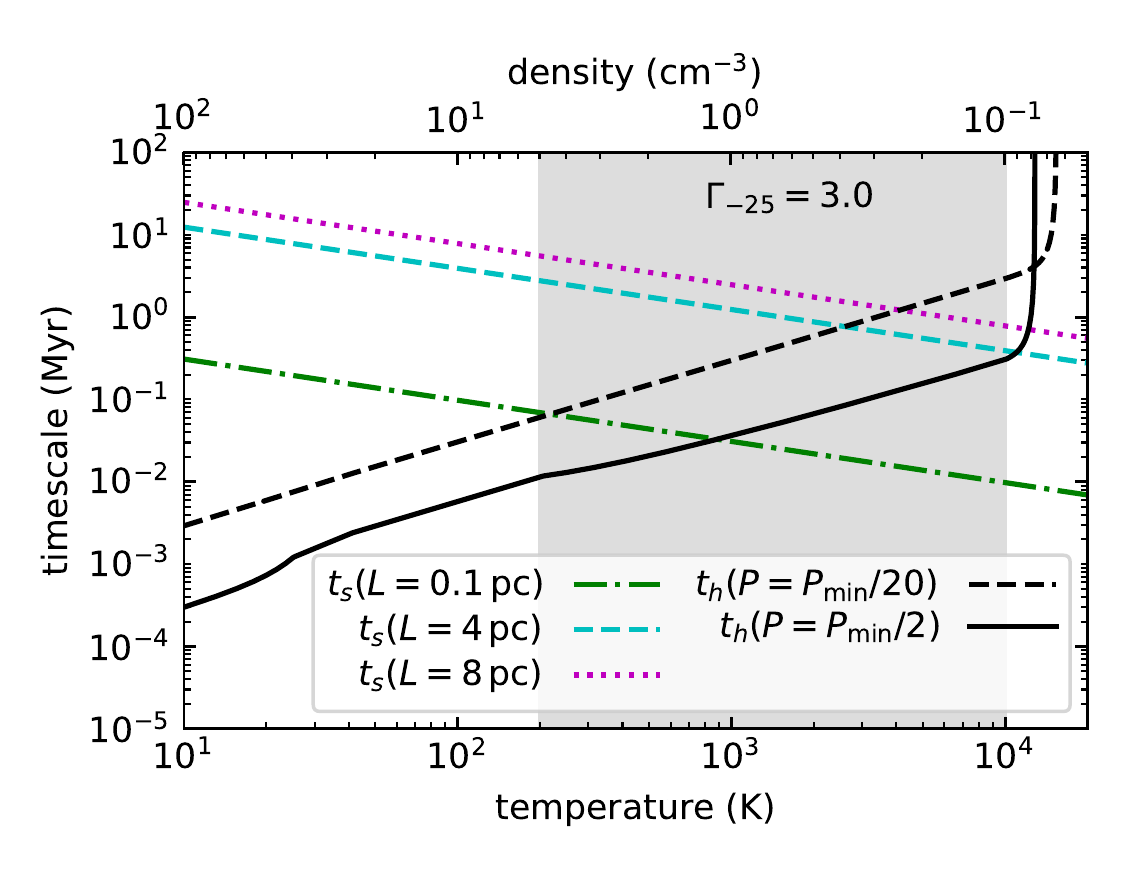}
\caption{Timescales for isochoric heating (eq.~[\ref{eq:t_heating}], black lines) and isobaric expansion ($t_s = L/c_s$, colored lines) of gas perturbed out of equilibrium with heating rate $\Gamma_{-25}=3$. The bottom axis shows the temperature; the top axis shows the equivalent density $n = \Pmin / 2kT$ for the solid black line.The thermally unstable regime (excluding the marginally-stable $39.8 \K < T < 200 \K$ range) is gray.}
\label{fig:timescales}
\end{figure}

In the WNM-only regime, turbulent pressure fluctuations can still drive regions of gas into the CNM phase. If the WNM is compressed isobarically to densities at which the gas is at CNM temperatures but $\Pth < \Pmin$, the gas is out of equilibrium. It can return to equilibrium by expanding isobarically (therefore heating) to return to the WNM branch or by heating isochorically to the CNM branch. Isobaric expansion would move the gas to the left on the pressure-density phase diagram; isochoric heating would move the gas upward on the phase diagram. Expansion of a cloud of size $L$ would occur on a sound crossing timescale, $t_s \sim L / c_s$ for sound speed $c_s$, whereas isochoric heating to the temperature where $\Pth = \Pmin$ would occur on a heating timescale,
\begin{equation} \label{eq:t_heating}
t_h \sim \frac{\Pmin/n}{\GPE - n \Lambda(T)}.
\end{equation}
We illustrate these timescales in Figure~\ref{fig:timescales} for our fiducial model. For gas parcels comparable to our resolution ($L = \Delta x_0 = 4 \pc$), the isochoric heating timescale is $t_h \lesssim 10 \kyr$, which is $\gtrsim 2$ orders of magnitude shorter than the sound crossing time, as noted by \citet{PiontekOstriker:2004}. If our resolution allowed compression to smaller sizes such as $L = 0.1 \pc$, the heating timescale would remain an order of magnitude shorter than the sound crossing time for thermally unstable cold gas at $\Pth = \Pmin/2 = 990 k \K \cucm$. Therefore, any WNM gas below \Pmin\ that is turbulently compressed to CNM densities is likely to wind up on the stable CNM branch with $\Pth \approx \Pmin$ rather than returning isobarically to the stable WNM branch. Once CNM gas exists at $\Pth \approx \Pmin$, any WNM gas with lower pressure would be underpressured, leading to expansion of the CNM and compression of the WNM to bring the phases into thermal pressure equilibrium at $\Pth \gtrsim \Pmin$. Consequently, even if the mean density of the warm and cold ISM is below $n_\mathrm{CNM}(\Pmin)$ --- placing the gas in the WNM-only regime --- this mechanism would increase the pressure in both phases to \Pmin.
At heating rates that are sufficiently high so that $\nbarwc < n_\mathrm{CNM}(\Pmin)$, conservation of mass requires that very little mass can be found in the dense CNM, as is found in our high-$\GPE$ simulations. However, this turbulent compression mechanism allows the CNM physics to set the thermal pressure of the WNM even when the CNM mass filling fraction is very small ($f_{m,c} < 10^{-3}$ in our {\tt pe1229} models), preventing the ISM from entering the WNM only ($\nbarwc < n_\mathrm{CNM}(\Pmin)$) regime.

The cold gas thermal pressure in our model may be underestimated due to resolution. Although the overall thermal pressure and volume and mass filling factors are well-converged between our $\Delta x_0 = 4 \pc$ and $2 \pc$ models, the scale height $h_z$ and density of the densest gas are not (Table~\ref{tbl:outputs}). Cold clouds are resolved by only a few zones, especially in models with smaller cold gas volume filling fractions (Figures~\ref{fig:images} and \ref{fig:images_pe300}). Therefore, at higher resolution, the cold gas would have the ability to compress further, increasing the cold gas thermal pressure. The difference in thermal pressure between the cold and warm gas in our simulations should thus be interpreted cautiously. This issue does not affect our conclusions in this subsection.

\begin{figure*}[tb]
\includegraphics[width=\textwidth]{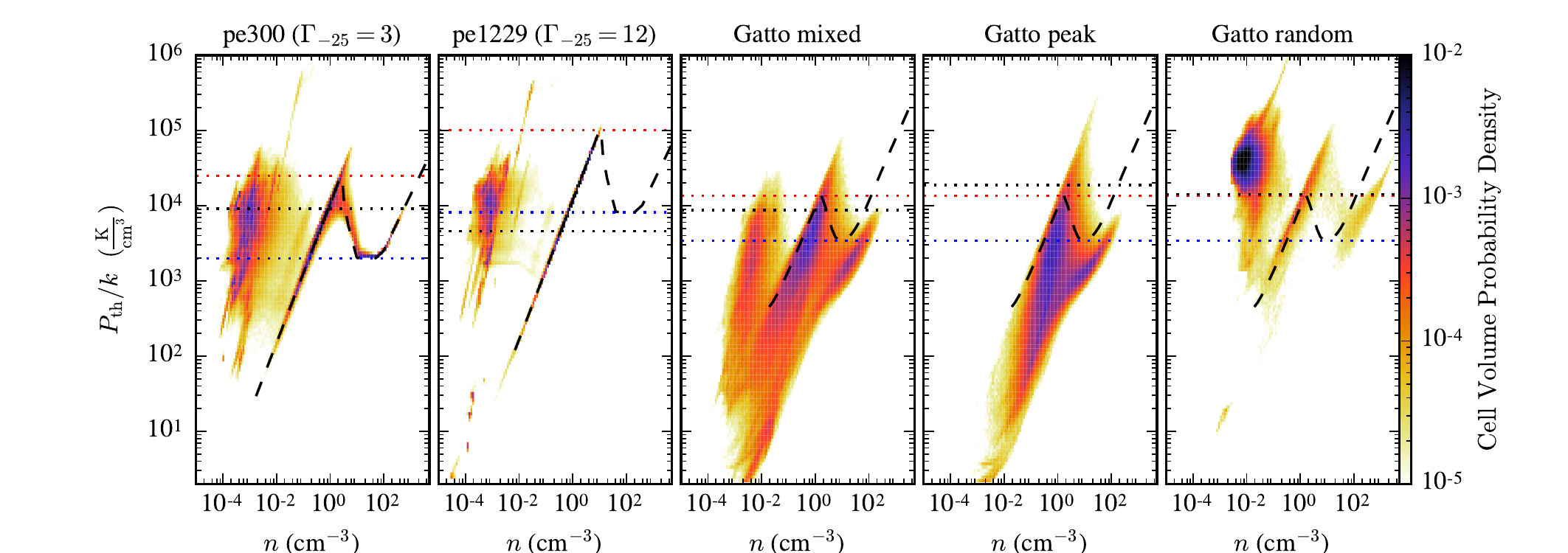}
\caption{Pressure-density phase diagrams for two of our fiducials models and three periodic box models from \citet{GattoWalch:2015}, all with $\Delta x_0 = 2 \pc$. The \citet{GattoWalch:2015} models use (left to right) mixed driving, peak driving, and random driving. As in Figure~\ref{fig:phase}, dotted red and blue lines show \Pmax\ and \Pmin, and dotted black lines show \Pturb\ calculated over the full box. The black dashed lines show the thrmal equilibrium curve.}
\label{fig:gatto_phase}
\end{figure*}

\subsection{Thermal pressure of hot gas in thermal runaway}

The HIM is mostly found in the range $\Pmin < \Pth < \Pmax$, even in our thermal runaway models ({\tt pe109}, the {\tt smd75} models, and {\tt nigm-4 bx0 pe300}). The hot gas is shock heated, a physical process not impacted by the balance between FUV heating and radiative cooling that determines the pressure of the two-phase CNM and WNM. Therefore, it must be thermal pressure equilibrium with the two-phase ISM that pushes the HIM towards the two-phase pressure regime.

For the periodic box models presented by \citet{GattoWalch:2015} and \citet{LiOstriker:2015}, the situation is qualitatively different. We show phase diagrams of three models from \citet{GattoWalch:2015} in Figure~\ref{fig:gatto_phase}. These models are similar to those presented in this paper except for a different cooling curve (leading to a narrower range between \Pmin\ and \Pmax), a supernova prescription that includes momentum as well as thermal energy injection, and fully periodic $256^3 \pc^3$ boxes without an external gravitational potential. These three models include peak driving, in which all supernovae are set off at density peaks; random driving, in which supernova locations are chosen randomly; and mixed driving, which is a mixture of the two.
In the peak driving case, the energy is injected in dense locations and thus radiates away quickly, resulting in the creation of no hot gas. In the random driving case, the simulations are in thermal runaway, with $92 \%$ of the volume in HIM gas and $96 \%$ of the mass in CNM gas. In the random driving case, the hot gas pressure is considerably higher than that in other phases, and $97\%$ of the hot gas has $\Pth > \Pmax$. The warm gas pressure is slightly lower, with the median $\Pth \approx \Pmax$, although the lowest-pressure gas found in the simulations is warm, as expected: only warm gas can be stable with $\Pth < \Pmin$. In the cold gas, the median thermal pressure is lower still, although only $\approx 2\%$ of the CNM has $\Pth < \Pmin$. As in our models, the cold clumps --- particularly in the thermal runaway case --- are at best only marginally resolved.

In the periodic box thermal runaway models (with random driving), the hot gas is overpressured, with $P_\mathrm{th,h} > \Pmax > P_\mathrm{th,c}$. We do not see this behavior in any of our models. This confirms the result of  \citet{WalchGirichidis:2015}, who use a similar code to \citet{GattoWalch:2015} but with boundary conditions allowing vertical stratification. Many of their models also have very high HIM filling fractions, particularly in the midplane (see their Fig.~25), yet their hot gas thermal pressure is, like our vertically-stratified models, in or near the stable two-phase pressure range. We conclude that once thermal runaway is reached, the lack of a pressure valve in fully-periodic models leads to unphysically high HIM thermal pressures.

\subsection{Caveats}
 \label{sec:missing}

Our focus in this paper is on a controlled physics experiment in which we vary the FUV heating rate. In doing so, we neglect a number of physical processes, including the following.

\begin{enumerate}
\item We use a $1 \kpc^2$ region of a galaxy in which there is little variation in the FUV radiation field either temporally or spatially. The FUV field likely varies by factors of $\sim 3$ on $\lesssim 100 \Myr$ timescales near the Sun \citep{ParravanoHollenbach:2003}. Over an entire galaxy, the radiation field decreases with galactocentric radius, reducing \Pmin. However, the value of \Pmin\ implied by the FUV radiation field as a function of Galactocentric radius as modelled by \citet{WolfireMcKee:2003} does not fall off quickly enough to explain the near-constant observed CNM-WNM ratio \citep{DickeyStrasser:2009}.
\item In reality, the {\em average} FUV radiation field is a function of the star formation rate, so varying the radiation field without varying the supernova rate is unphysical. However, the average radiation field may not apply in many places due to its wide spatial and temporal variance. Incorporating radiative transfer to fully model the time-variable FUV radiation field from individual B stars in a large set of MHD simulations like ours is prohibitively expensive. Approximations in which the box-wide FUV radiation field varies with the current star formation rate are feasible \citep{KimOstriker:2017a}, but there remains uncertainty in the FUV radiation field at any point as a function of the average star formation rate.
\item The efficiency of photoelectric heating depends upon the charge, size distribution, and composition of the dust grains \citep{Spitzer:1948,BakesTielens:1994,WeingartnerDraine:2001}. However, the local variations due to these factors are 
      unlikely to negate the qualitative picture presented here.
\item We use a cooling curve that approximates cooling due to solar neighborhood metallicity gas neglecting depletion with a constant ionization fraction of $10^{-2}$. This somewhat overestimates the gas-phase coolant abundances in the ISM near the Sun. If we included a cooling curve that accounted for depletion, cooling would likely be less efficient, effectively changing the shape and position of the thermal equilibrium curve, but not the basic relation between heating and expansion timescales.
\item Photoionization heating \citep{BarnesWood:2014,VandenbrouckeWood:2018}, cosmic ray heating \citep{WienerZweibel:2013,GirichidisWalch:2016}, and non-equilibrium ionization \citep{AvillezAsgekar:2012} are each likely to increase the scale height of the warm gas, resulting in additional midplane pressure.
\item We do not include gas self-gravity. Self-gravity can allow dense gas to decouple from thermal pressure equilibrium, allowing it to reach considerably higher pressure than the CNM or WNM, but is not important in the CNM, WNM, or WIM itself.
\item In all models, we chose the location and timing of supernovae without knowledge of the gas density. This is somewhat unphysical, although we do not resolve star formation in any meaningful way in these models and the B stars that produce most supernovae even in massive star clusters live longer than the giant molecular clouds in which they form. The major drawback of this piece of missing physics is that cold clouds are unphysically long-lived (with individual cold clouds surviving for $\gtrsim 100 \Myr$) due to the lack of internal feedback.
\item Because we do not include shearing boundary conditions, the magnetic field approaches an equilibrium considerably weaker than that in the real ISM \citep{GresselElstner:2008,HillJoung:2012,KimOstriker:2017a}.
\end{enumerate}

\subsection{Subgrid Models}
Many of the effects identified in Section~\ref{sec:missing} will quantitatively
modify the midplane pressure. However, we do not expect qualitative changes that
affect our main result: 
{\em that the relative timescales of isobaric expansion and isochoric heating of thermally-unstable gas with} $\Pth < \Pmin$ {\em tends to drive the WNM to} $\Pth \gtrsim \Pmin$. This microphysical process should occur regardless of the physical processes that set \Pmin\ and could in principle be applied as a minimum warm gas pressure in simulations that do not resolve the CNM. In applying this effect to such models, in which the CNM is unresolved, \Pmin\ should be chosen based on the appropriate, spatially-varying heating and cooling functions.

\section{Summary} \label{sec:summary}

In this paper, we conducted a set of numerical simulations of the diffuse ISM in a $1 \times 1 \times \pm 20 \kpc^3$ box. We included supernova feedback with a rate we controlled and positions independent of the gas density. We used a piecewise-power law cooling curve approximating the ISM conditions in the solar neighborhood and a diffuse heating term \GPE\ which approximates the FUV heating of interstellar dust grains. The primary experiment in this paper was to study the effects of varying \GPE. We also ran experiments with a magnetic field and different supernova rates, surface mass densities, and IGM densities. Our main conclusions follow.

\begin{enumerate}
\item The thermal pressure range $\Pmin < \Pth < \Pmax$ in which the two-phase ISM (CNM and WNM) can exist is set by the heating rate and therefore varies by more than an order of magnitude across our simulations, yet in every case, both CNM and WNM gas lie nearly entirely within the two-phase pressure range (Fig.~\ref{fig:p_gamma}).

\item 
The relative timescales for radiative heating and isobaric expansion provide the physical explanation for the tendency for the gas to remain in the two-phase pressure regime \citep{WolfireMcKee:2003,PiontekOstriker:2004}. If any WNM gas has $\Pth < \Pmin$, turbulent compression can create thermally unstable CNM pockets. For this unstable gas, the timescale to heat to \Pmin\ (eq.~[\ref{eq:t_heating}]) is much shorter than the timescale to expand isobarically, so turbulent compression followed by radiative heating brings the gas pressure up to \Pmin\ (Section~\ref{sec:stability}). This tendency to remain within $\Pmin < \Pth < \Pmax$ is set by the gas microphysics \citep{WolfireMcKee:2003} and is independent of any variations in the star formation rate that would occur in a self-regulation scenario.

\item The HIM also lies in the two-phase pressure range. In the low $\GPE$ models, the HIM thermal pressure is larger than the other two phases, whereas in the high $\GPE$ models, the HIM pressure is comparable to the CNM pressure (Fig.~\ref{fig:p_gamma}).

\item When models with periodic boundary conditions enter thermal runaway, the HIM is overpressured because without a vertical gravitational potential, the hot gas has no ability to expand upwards to relieve the the excess pressure and the hot gas cooling time is longer than the time before subsequent supernova shocks. In models with a vertical gravitational potential, the HIM can stratify vertically so the midplane HIM pressure is within the two-phase range even in models where the HIM is dominant (Fig.~\ref{fig:gatto_phase}).

\item At low heating rates, where the two-phase ISM is at relatively low pressure, the HIM is the dominant phase by volume (near the midplane) and the CNM is the dominant phase by mass. At high heating rates, the WNM is the dominant phase by both mass and volume. At intermediate heating rates ($\Gamma_0 \approx 3 \times 10^{-25} \erg \s^{-1}$ for our cooling curve and our unmagnetized setup with our fiducial supernova rate and surface mass density), the WNM and HIM each fill approximately half of the midplane volume while the WNM and CNM each account for approximately half of the mass (Fig.~\ref{fig:stats}).

\item The weight, $W$, of the ISM varies little with $\Gamma_0$. Because of the changing value of \Pmax, $W > \Pmax$ for the low-$\Gamma_0$ models but $W < \Pmax$ for intermediate- and high-$\Gamma_0$ models (Fig.~\ref{fig:p_gamma}).

\item The turbulent pressure adjusts to support the weight given the thermal pressure. When we calculate the turbulent pressure on large scales ($\gtrsim 500 \pc$, half the horizontal extent of our box), the sum of the turbulent and thermal pressures are approximately equal to the weight in all cases which are in hydrostatic equilibrium. When turbulent pressure is calculated on small scales or as $\langle \rho v_z^2 \rangle$, the total pressure is insufficient to provide hydrostatic support (Fig.~\ref{fig:pth_pturb}).

\item In developing sub-grid models in which the structure of the two-phase medium is not resolved, this work suggests that the thermal pressure of all three phases should be controlled to remain within the two-phase pressure regime given the expected heating rate.
\end{enumerate}

The data presented in this paper are available at the American Museum of Natural History Research Library Digital Repository at\dataset[doi:10.5531/sd.astro.2]{\doi{10.5531/sd.astro.2}}.

\acknowledgements

A.S.H.\ and M.-M.M.L.\ acknowledge support by NASA through grant number HST-AR-14297 from Space Telescope Science Institute, which is operated by AURA, Inc. under NASA contract NAS 5-26555, and by NASA ATP grant NNX17AH80G. A.S.H.\ was partially supported by NSF grant AST-1442650. J.C.I.-M.\ acknowledges funding by the Deutsche Forschungsgemeinschaft (DFG) via the Sonderforschungsbereich SFB 956 ``Conditions and Impact of Star Formation'' (subproject C5), and the support by the DFG Priority Program 1573 ``The physics of the interstellar medium''. Resources supporting this work were provided by the NASA High-End Computing (HEC) Program through the NASA Advanced Supercomputing (NAS) Division at Ames Research Center. We thank the anonymous referee for comments which led to the addition of Section~\ref{sec:missing}.

\software{Flash 4.2 \citep{FryxellOlson:2000}, yt \citep{TurkSmith:2011}}

\bibliography{bibdesk_bibtex}

\end{document}